\documentclass{aa}  

\usepackage{graphicx}
\usepackage{txfonts}
\newcommand{\teff}{\ifmmode T_{\rm eff} \else $T_{\mathrm{eff}}$\fi}
\newcommand{\logg}{\ifmmode \log g \else $\log g$\fi}
\newcommand{\lL}{\ifmmode \log \frac{L}{L_{\odot}} \else $\log \frac{L}{L_{\odot}}$\fi}
\newcommand{\mdot}{\ifmmode \dot{M} \else $\dot{M}$\fi}
\newcommand{\myr}{M$_{\odot}$ yr$^{-1}$}
\newcommand{\vsini}{\ifmmode v \sin i \else $v \sin i$\fi}
\newcommand{\vinf}{$v_{\infty}$}

\newcommand{\vmac}{\ifmmode v_{\rm mac} \else $v_{\rm mac}$\fi}

\newcommand{\kms}{km~s$^{-1}$}
\newcommand{\msun}{\ifmmode M_{\odot} \else $M_{\odot}$\fi}
\newcommand{\zsun}{\ifmmode Z_{\odot} \else $Z_{\odot}$\fi}
\newcommand{\lsun}{\ifmmode L_{\odot} \else $L_{\odot}$\fi}
\newcommand{\rsun}{\ifmmode R_{\odot} \else $R_{\odot}$\fi}
\newcommand{\qh}{\ifmmode Q_{\rm H} \else $Q_{\rm H}$\fi}
\newcommand{\qhei}{\ifmmode Q_{\ion{He}{i}} \else $Q_{\ion{He}{i}}$\fi}

\begin{document}

   \title{A modern study of HD\,166734: a massive supergiant system \thanks{Based on observations collected at the European Southern Observatory (La Silla, Chile) with FEROS and TAROT and on data collected at the San Pedro M\'artir observatory (Mexico).}}

   \author{L. Mahy\inst{1,5}\fnmsep\thanks{F.R.S.-FNRS Postdoctoral researcher}
          \and
          Y. Damerdji\inst{2,1}
          \and
          E. Gosset\inst{1}\fnmsep\thanks{F.R.S.-FNRS Senior Research Associate}
          \and
          C. Nitschelm\inst{3}
          \and
          P. Eenens\inst{4}
          \and
          H. Sana\inst{5}
          \and
          A. Klotz\inst{6}
          }

   \offprints{L. Mahy}

   \institute{
     Space sciences, Technologies, and Astrophysics Research (STAR) Institute, Universit{\'e} de Li{\`e}ge, Quartier Agora, B{\^a}t B5c, All{\'e}e du 6 ao{\^u}t, 19c, B-4000 Li{\`e}ge, Belgium\\
     \email{mahy@astro.ulg.ac.be}
     \and
     Centre de Recherche en Astronomie, Astrophysique et G\'eophysique, route de l’Observatoire BP 63 Bouzareah, 16340 Algiers, Algeria
     \and
     Unidad de Astronom{\'i}a, Facultad de Ciencias B{\'a}sicas, Universidad de Antofagasta, Antofagasta, Chile
     \and
     Departamento de Astronom\'ia, Universidad de Guanajuato, Apartado 144, 36000 Guanajuato, GTO, Mexico
     \and
     Instituut voor Sterrenkunde, KU Leuven, Celestijnenlaan 200D, Bus 2401, B-3001 Leuven, Belgium
     \and 
     Universit\'e de Toulouse, UPS-OMP, IRAP, Toulouse, France
     }

   \date{Received September 15, 1996; accepted March 16, 1997}

 
  \abstract
   {}
   {HD\,166734 is an eccentric eclipsing binary system composed of two supergiant O-type stars, orbiting with a 34.5-day period. In this rare configuration for such stars, the two objects mainly evolve independently, following single-star evolution so far. This system provides a chance to study the individual parameters of two supergiant massive stars and to derive their real masses.}
   {An intensive monitoring was dedicated to HD\,166734. We analyzed mid- and high-resolution optical spectra to constrain the orbital parameters of this system. We also studied its light curve for the first time, obtained in the {\it VRI} filters. Finally, we disentangled the spectra of the two stars and modeled them with the CMFGEN atmosphere code in order to determine the individual physical parameters. }
   {HD\,166734 is a O7.5If$+$O9I(f) binary. We confirm its orbital period but we revise the other orbital parameters. In comparison to what we found in the literature, the system is more eccentric and, now, the hottest and the most luminous component is also the most massive one. The light curve exhibits only one eclipse and its analysis indicates an inclination of $63.0\degr \pm 2.7\degr$. The photometric analysis provides us with a good estimation of the luminosities of the stars, and therefore their exact positions in the Hertzsprung-Russell diagram. The evolutionary and the spectroscopic masses show good agreement with the dynamical masses of 39.5\,\msun\ for the primary and 33.5\,\msun\ for the secondary, within the uncertainties. The two components are both enriched in helium and in nitrogen and depleted in carbon. In addition, the primary also shows a depletion in oxygen. Their surface abundances are however not different from those derived from single supergiant stars, yielding, for both components, an evolution similar to that of single stars. }
   {}

   \keywords{Stars: early-type - Stars: binaries: general - Stars: fundamental parameters - Stars: individual: HD\,166734}

   \maketitle
%

   \section{Introduction}
   Massive stars constitute a small part of the stellar population but have a considerable influence on their environment. About 70\% of them \citep{sana12}, and even 90\% if we consider orbital periods longer than 3500 days \citep{sana14}, are moreover expected to be binaries or multiple systems. This multiplicity can affect the way that they evolve through tidally-induced-rotational mixing, exchange of matter and angular momentum through Roche lobe overflow, or even the coalescence of both components. In the case of young systems with large separations, the different components evolve independently, following single-star evolution \citep{pavlovski05} whilst, when the components interact with each other, their evolution can completely differ from that of single stars, affecting their size, mass, surface chemistry and/or even their rotational rate.

   To better understand how massive stars evolve and improve evolutionary models, comparisons between the latter and spectroscopic observations are essential. Nevertheless, for single stars, the modeling by atmosphere codes does not provide all the information about their fundamental properties, especially on the real masses of the stars. \citet{herrero92} raised the attention to the mass discrepancy problem. This term refers to a disagreement between the mass determined by comparing the position of the star in the Hertzsprung-Russell (HR) diagram according to the predictions of stellar evolution calculations and the mass obtained from the spectroscopically derived values of the surface gravity and of the radius of the star. In this context, detached eclipsing binary systems are of crucial importance to provide direct measurements of the physical properties of stars such as their radius or their dynamical mass, allowing us to test the predictive abilities of evolutionary models for single stars.

   HD\,166734 ($=\,$\object{V411\,Ser}) is particularly promising for such an analysis. Unfortunately, this system has neither a measured inclination nor a cluster membership. It was located by \citet{conti80} at a distance of 2.3~kpc from Earth. It is composed of two supergiant O-type stars in a wide eccentric orbit. This configuration, rather rare for such stars, suggests few interactions between each other, making them perfect objects for testing the theoretical schemes of stellar evolution. HD\,166734 has not been investigated since the seventies when \citet{conti80} provided for the first time the parameters of its orbit. These authors described this system as an O7.5If primary and an O9I secondary moving on an eccentric orbit ($e=0.46$) with a period of 34.54\,days. These two objects had similar luminosities but the secondary presents a slightly larger minimum mass than the primary (31\,\msun\ vs 29\,\msun, respectively). On the basis of these minimum masses \citet{conti80} predicted eclipses in the light curve of HD\,166734. Even though \citet{otero05} mentioned HD\,166734 as photometrically variable, the light curve has never been subject to a dedicated analysis. 

   The present paper aims at showing the advances that have been achieved in the analysis of this interesting system and is organized  as follows. The spectroscopic and photometric data are described in Section\,\ref{sec:obs}, together with their reduction procedure. The radial velocity (RV) determination and the orbital solution of HD\,166734 are presented in Section\,\ref{sec:RVorb}. Section\,\ref{photometry} reports the analysis of the {\it VRI} light curves. In Section\,\ref{sec:modeling}, we disentangle the spectra of each component and we model them with an atmosphere code to constrain the individual parameters and the surface abundances of each star. Section\,\ref{sec:wind} is devoted to the winds of both components and to their interactions. Section\,\ref{sec:discussion} provides the evolutionary statuses of both components and discusses the distance to HD\,166734. Finally our conclusions are given in Section\,\ref{sec:conclusion}.
   

   \section{Observations and data reduction}
   \label{sec:obs}
   \subsection{Spectroscopy}
   HD\,166734 was observed through a multi-epoch campaign with two different instruments. We first acquired high-resolution optical spectra with FEROS (Fiber-fed Extended Range Optical Spectrograph) successively mounted on the ESO-1.52m (for the observations taken before 2002) and on the ESO/MPG-2.2m telescopes at La Silla (Chile). This instrument has a resolving power of about 48\,000 and the detector was a 2k $\times$ 4k EEV CCD with a pixel size of 15$\mu$m $\times$ 15$\mu$m. It provides \'echelle spectra composed of 39 orders, allowing us to cover the [3800--9200]\AA\ wavelength domain. The data reduction was performed with an improved version of the FEROS pipeline as described by \citet{sana06}. The data normalization was performed by fitting polynomials of degree 4 or 5 to carefully chosen continuum windows.

   We completed this dataset by acquiring medium-resolution \'echelle spectra with the ESPRESSO spectrograph. This instrument is mounted on the 2.12m telescope at Observatorio Astron\'omico Nacional of San Pedro M\'artir (SPM) in Mexico. The spectra are slitted over 27 orders to cover the [3800--6950]\AA\ region. The spectral resolving power is of about 18\,000. To increase the signal-to-noise ratio of our data, we took several consecutive observations and we combined them. The data were reduced by using the \'echelle package available within the ESO-MIDAS software.

   The journal of the observations is provided in Table\,\ref{tab:journal} in the Appendix.

   \subsection{Photometry}

   We obtained the light curve of HD\,166734 in the {\it VRI} filters with the TAROT (T\'elescope \`a Action Rapide pour l'Observation des ph\'enom\`enes Transitoires) robotic observatory located at La Silla. It is a fully autonomous 25\,cm aperture telescope ($F/D = 3.4$) covering a $1.86\degr \times 1.86\degr$ field of view thanks to an Andor CCD camera. Its spatial sampling is $3.3\arcsec$/pixel. The reduction method of the TAROT observations is described in \citet{damerdji07}. The object was observed between March 2011 and September 2016 in the {\it V} filter, between February 2010 and September 2016 in the {\it R} filter and between February 2011 and September 2016 in the {\it I} filter. These three light curves are composed of 5021, 5178, and 4928 points, respectively.
   
   
   \section{Radial velocity measurements and orbital solution}
   \label{sec:RVorb}
\begin{figure}
     \centering
     \includegraphics[width=9cm,bb=23 1 526 398,clip]{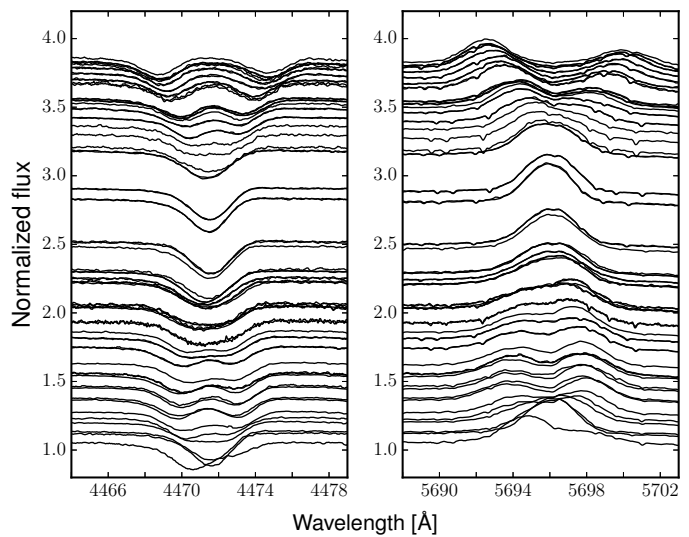}
     \caption{FEROS spectra in the region of \ion{He}{i}~4471 (left) and \ion{C}{iii}~5696 (right) shifted vertically as a function of phase (times 3 for clarity) from 0.0 at the bottom to 1.0 at the top.}\label{fig:spectra} 
   \end{figure}

   Figure\,\ref{fig:spectra} shows FEROS spectra as a function of the orbital phase (computed from orbital parameters determined below and listed in Table\,\ref{tab:orbital}). It illustrates the variability of the line profiles for two particular regions featuring the \ion{He}{i}~4471 absorption line as well as the \ion{C}{iii}~5696 emission line.

   To construct the orbital solution of HD\,166734, we first measure the RVs of both components. For this purpose, we fit their spectral lines with Gaussian profiles. The computations of the RVs are done by adopting the rest wavelengths from \citet{con77} for the lines with a rest wavelength shorter than 5000~\AA, and those of \citet{und94} for the others. Then we refine those values (notably for the spectra where the lines of both components are blended) by using a new method designed by one of the authors (YD). The latter is based on a minimum distance method considering a comparison between the observed spectrum and a composite (two-components) theoretical template. We perform this measurement for several lines individually, no matter their ionization stage. The final set of RVs (provided in Table\,\ref{tab:journal} in the Appendix) is obtained by computing the mean values of the radial velocities measured on all the various spectral lines. 

   \begin{table}
     \begin{center}
       \caption{Orbital solution of HD\,166734. The given errors correspond to 1~-~$\sigma$.} \label{tab:orbital}
       \begin{tabular}{lrr}
         \hline\hline
         & \multicolumn{2}{c}{SB2 solution}\\
         & Primary        & Secondary \\
         \hline
         $P$~[day] & \multicolumn{2}{c}{34.537723 $\pm$ 0.001330} \\
         $e$ &  \multicolumn{2}{c}{$0.618 \pm 0.005$}\\
         $\omega$~[$\degr$] &  \multicolumn{2}{c}{$236.183 \pm 0.786$}\\
         $T_0$~[HJD~$-$~2\,450\,000]  & \multicolumn{2}{c}{2195.064 $\pm$ 0.036}\\
         $q~(M_1/M_2)$ & \multicolumn{2}{c}{1.179 $\pm$ 0.016}\\
         $\gamma$~[\kms] & $-5.88 \pm 0.89$ & 7.12 $\pm$ 0.96 \\
         $K$~[\kms] &  142.01 $\pm$ 1.53 & 167.39 $\pm$ 1.79 \\
         $a \sin i$~[$R_{\odot}$] & 76.23 $\pm$ 0.89 & 89.85 $\pm$ 1.05 \\
         $M \sin^3 i$~[\msun] & 27.92 $\pm$ 0.81 & 23.69 $\pm$ 0.68 \\
         rms~[\kms] & \multicolumn{2}{c}{$11.14$}\\
         \hline
       \end{tabular}
       \tablefoot{
         The orbital solution is computed from 129 RV measurements (listed in the Appendix). $T_0$ refers to the periastron passage. $\gamma$, $K$ and $a \sin i$ denote the apparent systemic velocity, the semi-amplitude of the radial velocity curve and the projected separation between the center of the star and the center of mass of the binary system.}
      \end{center}
   \end{table}
   
\begin{figure}
     \centering
     \includegraphics[width=9cm,bb=11 5 526 398,clip]{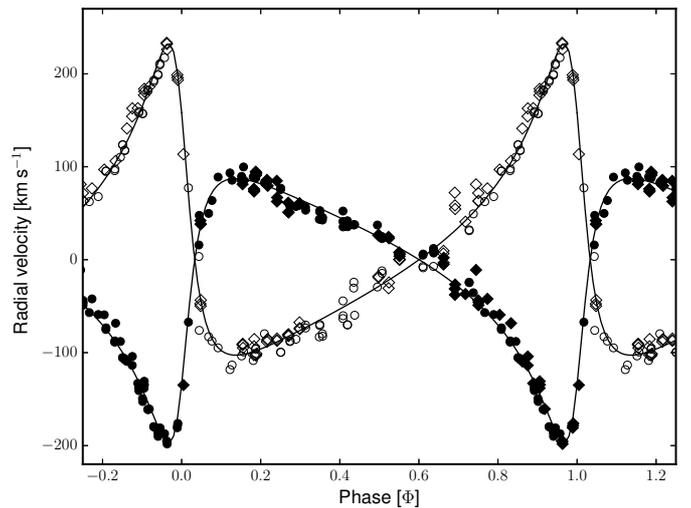}
     \caption{Radial velocity curves of HD\,166734 computed with the RVs from the present paper (provided electronically). The primary is represented by filled symbols whilst the open ones indicate the RVs of the secondary. Circles represent the RVs measured on the FEROS spectra and diamonds indicate the RVs measured on the ESPRESSO data. $\Phi = 0.0$ corresponds to the periastron passage.}\label{fig:rvcurve} 
   \end{figure}

   The determination of the orbital period is performed from the Heck-Manfroid-Mersch technique \citep[hereafter HMM,][]{heck85} that was revised by \citet[][see their appendix]{gosset01}. This technique is basically a Fourier method generalized to address the unevenly sampled time series. The HMM power spectrum is determined independently of the mean of the data and thus does not suffer from the biases of the \citet{scargle82} method. As input of this technique, we first use the RVs measured on the \ion{N}{iv}~4058 line that is only visible in the primary's spectrum. The periodogram exhibits two outstanding peaks at $\nu = 0.028453$~d$^{-1}$ and at $\nu = 0.028890$~d$^{-1}$ with a strong neighbor at $\nu = 0.028981$~d$^{-1}$. However, given their large width, the two last-quoted frequencies overlap. We then use HMM on our final set of RVs (provided in Table\,\ref{tab:journal}), on the basis of the difference between the secondary RVs and the primary RVs (RV$_{\mathrm{S}}$--RV$_{\mathrm{P}}$), and detect a frequency at $\nu = 0.028890$~d$^{-1}$, giving the period of $34.614053$~days and a neighbor at $\nu = 0.028957$~d$^{-1}$, corresponding to a period of 34.534560~days. We determine an error of $\epsilon(f) = 1.1\,10^{-6}$~d$^{-1}$ (corresponding to an error on the period of $1.3\,10^{-3}$~days) calculated on our dataset from the expression given by \citet[][see also \citealt{montgomery99} or \citealt{mahy11}, for further explanations]{lucy71}. We compare these values to the result obtained by \citet{conti80}. We stress that these authors determined an orbital frequency of $\nu = 0.028952$~d$^{-1}$ from the difference between the RVs of the two stars, but that the computed periodogram also shows a neighbor at $\nu \sim 0.02896$~d$^{-1}$ that overlaps with the other peak. To decrease the widths of the peaks, we thus need a dataset covering a longer timescale. We merge the two datasets and we determine in the periodogram once again two outstanding peaks at $\nu = 0.028879$~d$^{-1}$, corresponding to a period of 34.627837~days and at $\nu = 0.028954$~d$^{-1}$, corresponding to a period of 34.537542~days. By coupling the dataset of \citet{conti80} with our dataset, the error on the frequency becomes $\epsilon(f) = 2.4\,10^{-7}$~d$^{-1}$ which corresponds to an error on the period of $2.8\,10^{-4}$~days. The periodogram is exhibited in Fig.\,\ref{fig:periodogramRV} in the Appendix. We then use the Li{\`e}ge Orbital Solution Package (LOSP\footnote{LOSP is developed and maintained by H. Sana and is available at http://www.stsci.edu/$\sim$hsana/losp.html. The algorithm is based on the generalization of the SB1 method of \citet{wol67} to the SB2 case along the lines described in \citet{rau00} and \citet{sana06}.}) to determine the SB2 orbital solution of the system. The refined value of the orbital period, provided by this program (in order to improve the $\chi^2$ of the fit) is $34.537723 \pm 0.001330$ days and is adopted in the following. We thus confirm the period computed by \citet{conti80} on the basis of the difference between their secondary RVs and their primary RVs and we emphasize that our refined value is similar to the period associated to the alias found in the periodogram computed from the two coupled datasets. The parameters are listed in Table~\ref{tab:orbital} whilst the RV curves are shown in Fig.~\ref{fig:rvcurve}. From these orbital parameters, the inferior conjunction is expected at $\Phi \sim 0.018$ and the superior conjunction is at $\Phi \sim 0.766$.

   Our orbital solution presents an orbit more eccentric (0.62 vs. 0.46, respectively) than that derived by \citet{conti80}. Moreover, our monitoring (more intensive near the periastron passage) allows us to better constrain the semi-amplitudes, and thus the minimal mass of each component. In this new orbital solution, the primary is the most massive component of the system, which was not the case in the analysis of \citet{conti80}. This result seems thus to suggest that the two components of HD\,166734 show a `more classical' evolution than was first suspected \citep{conti80}.

   
   \section{Photometry}
   \label{photometry}
\begin{figure}
     \centering
     \includegraphics[width=9cm,bb=0 5 568 425,clip]{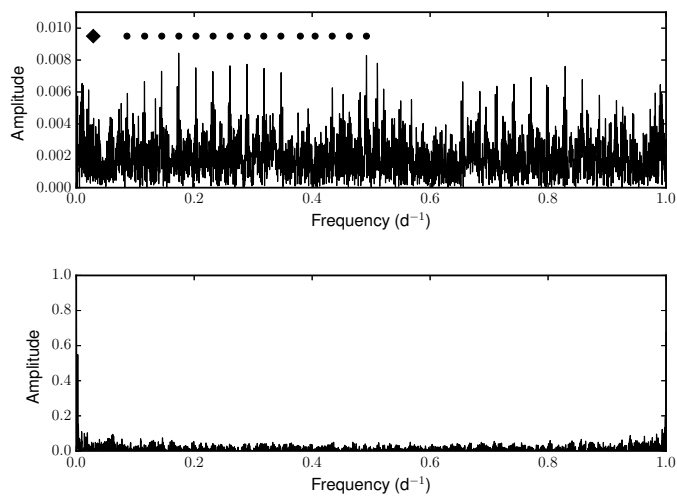}
     \caption{{\it Top:} Power spectrum of the {\it V} light curve of HD\,166734. The diamond indicates the location of the orbital frequency while the dots represent its harmonics. {\it Bottom:} Spectral window computed from the {\it V} light curve. }\label{fig:periodogramV} 
   \end{figure}

   Before studying the light curve to constrain the parameters of both objects, we submit the three photometric datasets of HD\,166734, obtained in the {\it VRI} filters, to a detailed Fourier analysis by HMM in order to search for the periods. Surprisingly, no outstanding peak is detected around 0.0289~d$^{-1}$ as would have been expected from the orbital frequency. Instead, we detect in each of the three datasets a series of peaks spaced with the same frequency of 0.0289~d$^{-1}$ (similar to the orbital frequency), starting at $\nu = 0.087$~d$^{-1}$ until $\nu = 0.492$~d$^{-1}$ (Fig.\,\ref{fig:periodogramV}). The dominant member varies from one filter to another, going from $\nu = 0.174$~d$^{-1}$ in the {\it V} and {\it R} filters to $\nu = 0.145$~d$^{-1}$ in the {\it I} filter (see Figs.\,\ref{fig:periodogramR}, \ref{fig:periodogramI} in the Appendix). This series of frequencies corresponds in fact to harmonics of the orbital frequency, even though the fundamental and the first-order harmonics are not detected. This large number of harmonics is explained by the small size of the eclipse with respect to the orbital cycle. Moreover, these harmonics have aliases between 0.5 and 1.0~d$^{-1}$ and at higher frequencies. We also apply HMM on the publicly available ASAS data and still find most of these harmonics there.

\begin{figure*}
     \centering
     \includegraphics[width=14cm,bb=26 184 582 602,clip]{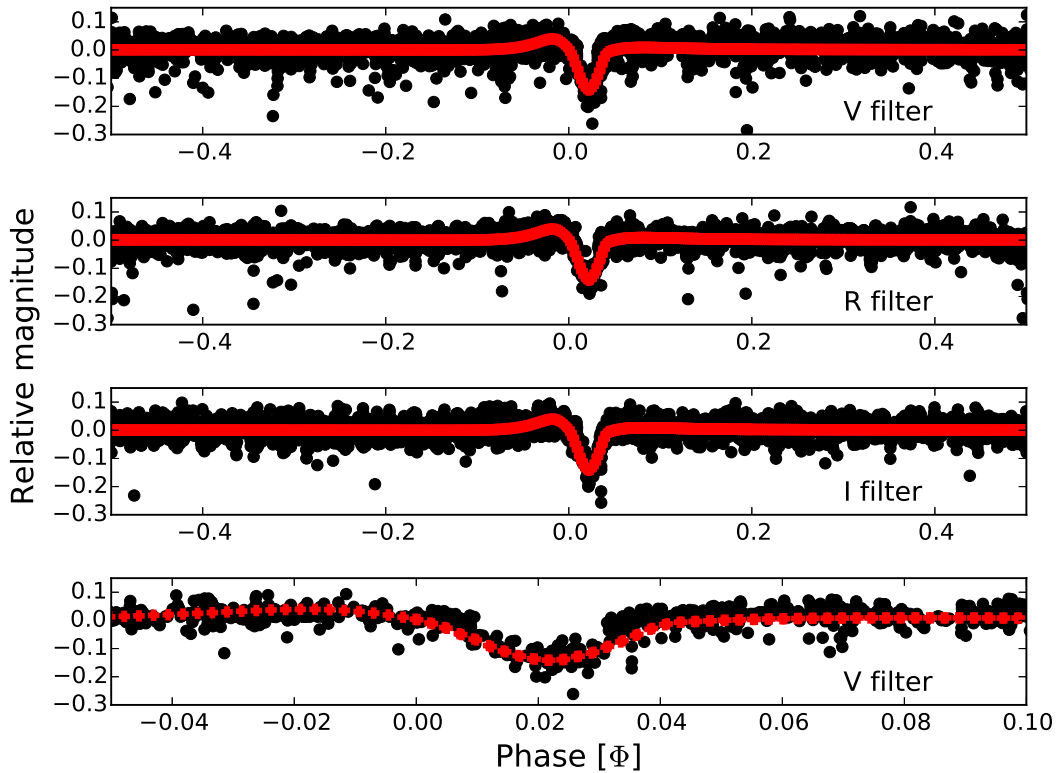}
     \caption{Light curves of HD\,166734 obtained with TAROT in the {\it V} (top panel), {\it R} (second panel) and {\it I} (third panel) filters. The red lines represent the best fit computed with PHOEBE. The magnitude outside the eclipse has been set to 0 in the three filters. The representation shown at the bottom is a zoom-in on the eclipse observed in the {\it V} light curve (plotted at the top of this Figure).}\label{fig:lightcurve} 
   \end{figure*}

   Because of the inclination of the orbit and the orientation of its apses with respect to our line of sight, the light curve of HD\,166734 exhibits only one eclipse during its orbital cycle, immediately after the periastron passage ($\Phi = 0.024$) with a total duration in orbital phase of $\Delta \Phi \sim 0.04$. The phase of this eclipse is in agreement with the estimation of \citet{otero05} but differs from the phase of the conjunction (see Section\,\ref{sec:RVorb}). This difference can come from the errors on the orbital parameters. We emphasize that if we fix the position of the eclipse to the conjunction it does not change the values of the orbital parameters within the uncertainties. This eclipse shows a difference of magnitude of about 0.22 mag and is preceeded by an increase of the total brightness of the system by about 0.03 mag at the periastron passage. We display in Fig.\,\ref{fig:lightcurve} the light curve of HD\,166734 in the three different filters sampled over the orbital period. We stress that these light curves contain outliers (with an unknown origin) that do not change their general shape. 

   To limit the possible effects of the degeneracies that can be produced because of the presence of only one eclipse we use the PHOEBE (PHysics Of Eclipsing BinariEs, v0.31a, \citealt{prsa05}) software while keeping the parameters of the orbit fixed at their values obtained through the spectroscopic analysis (Section\,\ref{sec:RVorb}, Table\,\ref{tab:orbital}). PHOEBE allows one to model the light curve and the RV curves of the object at the same time. This software is based on Wilson \& Devinney's code \citep{wilson71} and uses Nelder \& Mead's Simplex fitting method to adjust all the input parameters and to find the best fit to the light curve. Furthermore, it allows one to model multiple bandpasses at once. We include the reflection effects between the two stars in the computation of the light curve and we also fix the effective temperatures to 32000\,K and 30500\,K, for the primary and the secondary components, respectively, as determined in Section\,\ref{sec:modeling}. The photometric parameters derived from the modeling of the light curve are given in Table\,\ref{tab:photometric}. The error bars are computed by exploring the parameter space. For this purpose, we fix one parameter and allow the others to vary in order to reach the minimum of the $\chi^2$ corresponding to a 68.3\% confidence level (1-$\sigma$). Fig.\,\ref{fig:lightcurve} displays the best-fit of the light curve obtained by PHOEBE in the three different bandpasses.

   The system is seen under an inclination of $63.0\degr \pm 2.7\degr$. The eclipse occurs when the primary is in front of the secondary (secondary eclipse) along the line of sight. The two components do not fill their Roche lobe over the entire orbit. They fill about 90\%, within an error of 4\%, at the periastron displaying a small deformation because of tidal interactions. With this inclination the masses obtained from the orbital solution are estimated to $39.5^{+5.4}_{-4.4}$\,\msun\ for the primary and $33.5^{+4.6}_{-3.7}$\,\msun\ for the secondary. Their radii are computed to $27.5 \pm 2.3$\,\rsun\ and $26.8\pm 2.4$\,\rsun, respectively.

   The secondary component thus appears slightly more evolved (cooler and about the same size) than the primary. The modeling of the light curve also reports a primary component brighter than the secondary. From their radii and their effective temperatures, we determine stellar luminosities of $\lL = 5.840 \pm 0.092$ for the primary and of $\lL = 5.732 \pm 0.104$ for the secondary, infering a brightness ratio between the two stars of $l_1/l_2 = 1.28 \pm 0.14$. 
   
\begin{table}
\caption{Photometric parameters of HD\,166734 obtained with PHOEBE. The uncertainties correspond to 1--$\sigma$. }  
\label{tab:photometric} 
\centering            
\begin{tabular}{l c c}  
  \hline\hline
  & Primary & Secondary \\
  \hline
  $i$ [$\degr$] & \multicolumn{2}{c}{$63.0 \pm 2.7$} \\
  $R_{\mathrm{mean}}$ [\rsun] & $27.5 \pm 2.3$ &  $26.8 \pm 2.4$ \\
  $R_{\mathrm{pole}}$ [\rsun] & $26.5 \pm 2.0$ &  $25.9 \pm 2.2$ \\
  $R_{\mathrm{point}}$ [\rsun] & $31.6 \pm 5.4$ &  $31.9 \pm 7.3$ \\
  $R_{\mathrm{side}}$ [\rsun] & $27.1 \pm 2.1$ &  $26.0 \pm 2.2$ \\
  $R_{\mathrm{back}}$ [\rsun] & $28.8 \pm 2.8$ &  $28.1 \pm 2.8$ \\
  $M_{\mathrm{bol}}$ [mag] & $-9.85 \pm 0.17$ & $-9.58 \pm 0.20$ \\
  $\Omega$ & 3.499 & 3.026 \\
  $A$ & 1.0 & 1.0 \\
  $g$ & 1.0 & 1.0 \\
  $l_1/l_2$ &  \multicolumn{2}{c}{$1.28\pm0.14$}\\
  \hline
\end{tabular}
\tablefoot{
  $R_{\mathrm{pole}}$ is the radius toward the pole, $R_{\mathrm{point}}$ is the radius measured toward the other component, $R_{\mathrm{side}}$ is the radius measured toward the direction perpendicular to the two above-mentioned axes, $R_{\mathrm{back}}$ toward the Lagrangian point $L_2$ and $R_{\mathrm{mean}}$ is the geometrical mean between $R_{\mathrm{pole}}$, $R_{\mathrm{side}}$ and $R_{\mathrm{back}}$. $\Omega$ refers to the value of the Roche-model potential used in the WD program. $A$ are the bolometric albedos (coefficients of reprocessing of the emission of a companion by "reflection") and $g$ represents the gravity darkening coefficients. }

\end{table}


\section{Individual parameters}
\label{sec:modeling}

\subsection{Disentangling}

We use the orbital parameters derived in Section\,\ref{sec:RVorb} as input of the disentangling program. We apply the Fourier approach of \citet{hadrava95} to separate the spectral contributions of both components. This method uses Nelder \& Mead's Downhill Simplex on the multidimensional parameter space to reach the best $\chi^2$ fit between the recombined component spectra and the observed data. We emphasize that we only apply this technique on the FEROS spectra because they have a better signal-to-noise ratio and a higher resolution.

HD\,166734 is a detached eclipsing binary. During the eclipse, the secondary component is however not totally eclipsed. There exists therefore an ambiguity in the determination of the continuum level. The method of Fourier presented by \citet{hadrava95} induces oscillations in the continuum of the disentangled spectrum of each component when one of them is not completely eclipsed, as is the case for HD\,166734. A renormalization of the disentangled spectra must thus be done with the light ratio derived from the light curve in Section\,\ref{photometry} and provided in Table\,\ref{tab:photometric}. A discussion of the spectral disentangling can be found in \citet{pavlovski10} and in \citet{pavlovski12}.

   \begin{figure*}
     \centering
     \includegraphics[width=14cm,bb=3 0 559 420,clip]{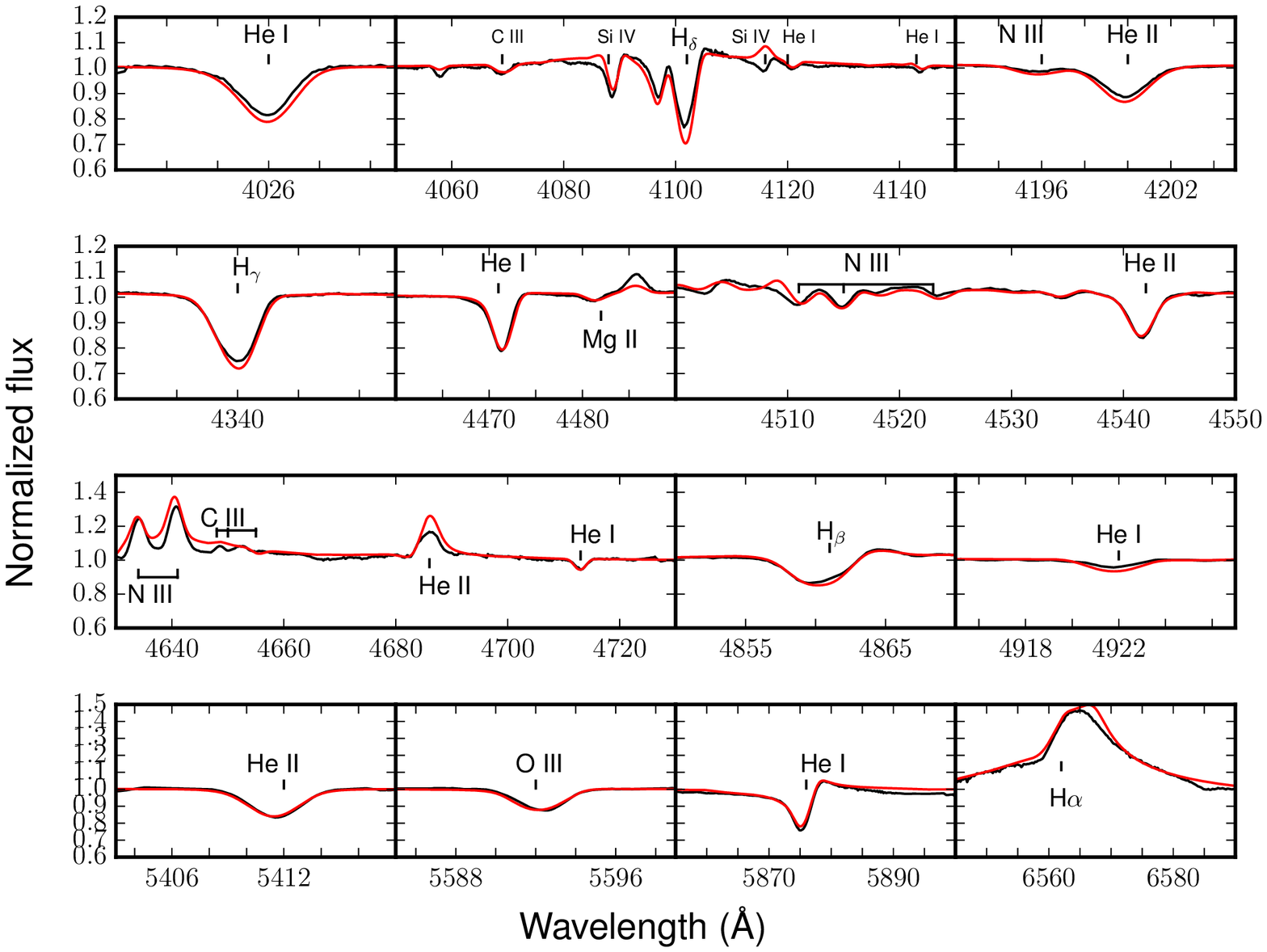}
     \caption{Best CMFGEN fit (red) of the disentangled optical spectrum (black) of the primary component of HD\,166734. }\label{primary_spectrum} 
   \end{figure*}
   
    To better constrain the continuum level of both disentangled spectra, we decided, however, to compare them with those obtained from the method of \citet{marchenko98}, revised by \citet{gonzalez06}. As already stressed by \citet{mahy15}, the technique of \citet{gonzalez06} gives reliable results outside broad lines such as the Balmer lines. In the wings of those lines, the disentangled spectra exhibit intrinsic excesses or deficiencies of the flux, making them difficult to normalize, and infering larger uncertainties on the shape of their wings. The two techniques have thus their respective advantages, and their comparison allows us to remove the artefacts linked to the disentangling method.

   \begin{figure*}
     \centering
     \includegraphics[width=14cm,bb=3 0 559 420,clip]{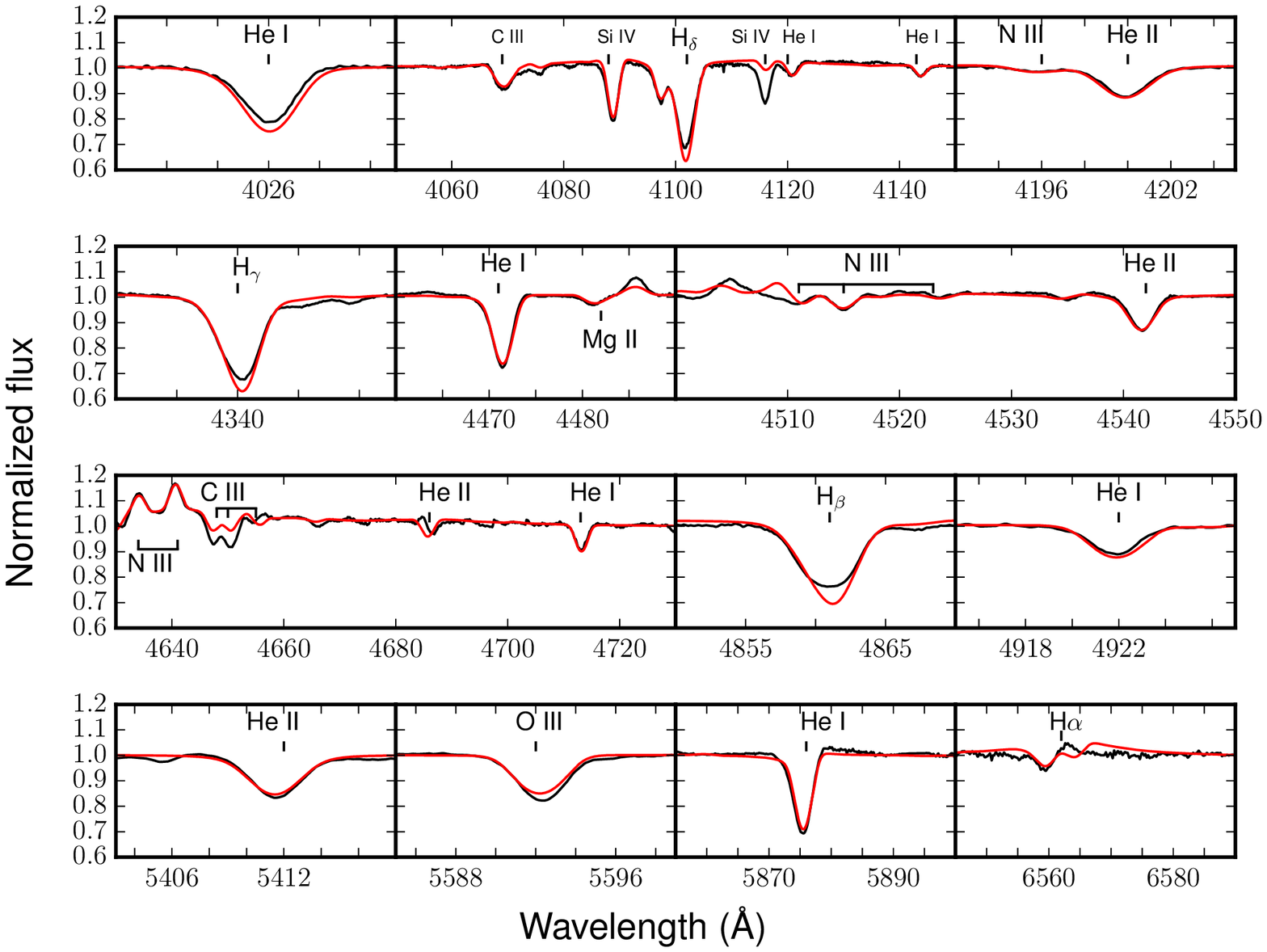}
     \caption{Same as Fig.\,\ref{primary_spectrum} but for the secondary component of HD\,166734.}\label{secondary_spectrum} 
   \end{figure*}
   
As mentioned above, the resulting spectra are normalized to remove any oscillation of the continuum linked to the Fourier disentangling technique and then corrected for the brightness ratio of $l_1/l_2 = 1.28$. The final disentangled primary and secondary spectra are displayed in Fig.\,\ref{primary_spectrum} and in Fig.\,\ref{secondary_spectrum}, respectively. 

   \subsection{Spectral-type classification}

   From the final disentangled spectra, we determine the equivalent widths of several lines to apply Conti's criteria \citep{conti71, conti73, mathys88, mathys89} and derive the spectral classifications of both components. We measure ratios of $\log (\ion{He}{i}~4471/\ion{He}{ii}~4542) = 0.06$ and $0.31$, which gives spectral classifications of O7.5 for the primary and of O9 for the secondary. We can then determine the luminosity class of each component using $\log (\ion{Si}{iv}~4089/\ion{He}{i}~4143)$. We find that this value is equal to 0.93 and to 0.65 for the primary and the secondary components, respectively, indicating that both stars are supergiant. We observe the lines of \ion{He}{ii}~4686 and \ion{N}{iii}~4634--41 in strong emission in the spectrum of the primary. According to \citet{walborn71}, we add the `f' suffix to the spectral type of the primary. For the secondary, we observe in its final disentangled spectrum that the absorption of the \ion{He}{ii}~4686 line is compensated by the emission, whilst the \ion{N}{iii}~4634--41 lines are moderately in emission, infering a `(f)' suffix to its spectral classification.
   
   We also compare the component spectra of both stars of the system to the spectral atlas of \citet{walborn90}, confirming their classification.
   
   In summary, we attribute to the primary an O7.5~If spectral type whilst the secondary is classified as an O9~I(f) star, with an uncertainty of one subtype for each component. These classifications are in agreement with those obtained by \citet{conti80}, within the uncertainties. 
   
   \subsection{Modeling}
   
   To determine the physical parameters of both components, we use the CMFGEN atmosphere code \citep{hillier98} on the disentangled spectra corrected for the brightness ratio. This code provides non-LTE atmosphere models including winds and line-blanketing. CMFGEN requires input from an estimate of the hydro-dynamical structure connected to a $\beta$ velocity law of the form $v = v_{\infty} (1-R/r)^{\beta}$, where $v_{\infty}$ is the terminal velocity of the wind. Our final models include the following chemical elements: \ion{H}{i}, \ion{He}{i-ii}, \ion{C}{ii-iv}, \ion{N}{ii-v}, \ion{O}{ii-v}, \ion{Al}{iii}, \ion{Ar}{iii-iv}, \ion{Mg}{ii}, \ion{Ne}{ii-iii}, \ion{S}{iii-iv}, \ion{Si}{ii-iv}, \ion{Fe}{ii-vi}, and \ion{Ni}{ii-v} with the solar composition of \citet{grevesse07} unless otherwise stated. CMFGEN allows a super-level approach to reduce the memory requirements. Therefore, we include about 1771 super levels for a total of about 8075 levels. The formal solution of the radiative transfer equation allows us to compute an emergent spectrum. To do so, we use a depth-independent microturbulent velocity varying linearly from 20\,\kms\ at the photosphere to $0.1 \times v_{\infty}$ at the top of the atmosphere. The choice of 20\,\kms\ for the microturbulence velocity at the photosphere is coherent with other studies made on supergiant massive star populations (see \citealt{bouret12} or \citealt{crowther09}). Once the emergent spectrum is calculated we convolve it twice, once with a rotational profile to include the projected rotational velocity (\vsini) and once with a Gaussian profile to include macroturbulence (\vmac) in the line profiles. \vsini\ is determined from the Fourier transform method given by \citet{simondiaz07} whilst \vmac\ is fixed by eye from the \ion{He}{i}~4713 and \ion{O}{iii}~5592 lines. We compute values of $ 95\pm 10$\,\kms\ and $65 \pm 10$\,\kms\ for the \vsini\ and the \vmac\ of the primary whilst values of $98 \pm 10$\,\kms\ and $65 \pm 10$\,\kms\ are obtained for the secondary, respectively.

   The effective temperature of each component is estimated from the ratio between \ion{He}{i} and \ion{He}{ii}, mainly from the \ion{He}{i}~4471 and \ion{He}{ii}~4542 lines, even though the other helium lines are also considered. The typical uncertainty on our determination is 1000\,K. The surface gravities of both components are determined from the wings of the Balmer lines H$\gamma$ and H$\delta$. The typical errors on the surface gravities of both components are 0.1~dex. H$\alpha$ and H$\beta$ are not taken into account because they are more affected by the stellar winds, or maybe even by the interaction between the winds of the two components. Furthermore, the shapes of these two lines can vary with respect to the orbital phase. In the final disentangled spectra, we obtain mean profiles computed over the entire orbital period that are not necessarily meaningful. The stellar luminosity is determined from the radii derived through the photometry (see Sect.\,\ref{photometry}) and from the effective temperature determined for each component through the CMFGEN best-fit (in an iterative way). The mass-loss rates and the clumping filling factors are adjusted to reproduce the H$\alpha$ and H$\beta$ lines as well as the \ion{He}{ii}~4686 line whilst the shape of their emission is sensitive to the $\beta$ exponent of the wind velocity law. The terminal velocity has been computed from the Newtonian escape velocity $v_{\mathrm{esc}} = \sqrt{2G\,M\,(1-\Gamma)/R}$ (where $G$ is the gravitational constant, $\Gamma$ the Eddington factor computed from \citet{vink01} by assuming an electron scattering cross-section per unit mass of 0.32 cm$^2$ g$^{-1}$ \citep{lamers93}, $M$ the mass of the star and $R$ its radius) by assuming $v_{\infty}/v_{\mathrm{esc}} = 2.6$.

   The best-fit model has been obtained by minimizing the calculated $\chi^2$ by varying different parameters in the parameter space. To this end, we generate a non-uniform grid composed of several dozen models for each star. Once all the fundamental parameters are constrained, we run models with different surface abundances (for He, C, N, and O). Specific lines of each element are selected from the disentangled spectra and we quantitatively compare these lines to the synthetic spectra by means of a $\chi^2$ analysis from which we derive the surface abundance and their uncertainties. This method is similar to that proposed by \citet{martins15}, except that, rather than renormalizing the $\chi^2$ function so that the minimum has a value of 1.0, we subtract an appropriate quantity to set the minimum to 1.0. The $1-\sigma$ uncertainty is approximately set to the abundances for which $\chi^2 = 1 + 1/\chi_{min}^2$. The lines used to constrain the abundances are \ion{C}{iii}~4068--4070, and \ion{C}{iii}~4153--56--63 for carbon, \ion{N}{iv}~4058 (only for the primary), and \ion{N}{iii}~4511--4515--4524 for nitrogen and \ion{O}{iii}~5592 for oxygen. We exclude the \ion{C}{iii}~4647--50--51, \ion{C}{iii}~5696, and \ion{C}{iv}~5801--5812 lines since their formation process depends on fine details of atomic physics and on the modeling \citep{martinshillier12}. Moreover, special care is taken for the helium abundance. Some of the helium lines are also affected by the microturbulent velocity taken to compute the emerging spectrum \citep{macerlean98}. A compromise is thus made to ensure the best CMFGEN fit. The individual parameters of both components are given in Table\,\ref{tab:parameter} and we display, in Figs.\,\ref{primary_spectrum} and\,\ref{secondary_spectrum}, the CMFGEN best-fit model for the primary and the secondary, respectively.

\begin{table}
\caption{Individual parameters of each component of HD\,166734}  
\label{tab:parameter} 
\centering            
\begin{tabular}{l c c }  
  \hline\hline
  & Primary & Secondary \\
  \hline
  \teff\ [K] & $32000 \pm 1000$ & $30500\pm 1000$ \\
  \logg\ [cgs] & $3.15 \pm 0.10$ & $3.10 \pm 0.10$ \\
  $\log (L/\lsun)$& $5.840 \pm 0.092$ &  $5.732 \pm 0.104$\\
  $\mdot / \sqrt{f}$ [\myr] & $9.07 \times~10^{-6}$ & $3.02 \times~10^{-6}$\\
  \vinf\ [\kms] & $1407$ &  $1348$ \\
  $f$ & 0.07 &  0.07 \\
  $v_{\mathrm{cl}}$ [\kms] & 30 & 30\\
  $\beta$  & 1.0 & 1.0\\
  He/H & $0.12 \pm 0.03$ & $0.12 \pm 0.03$ \\
  C/H & $1.2 \pm 0.4 \times~10^{-4}$ & $2.0 \pm 0.3 \times~10^{-4}$  \\
  N/H & $6.1 \pm 1.2 \times~10^{-4}$ & $1.8 \pm 0.5 \times~10^{-4}$ \\
  O/H & $2.7 \pm 0.5 \times~10^{-4}$ & $4.6 \pm 0.3 \times~10^{-4}$ \\
  \vsini\ [\kms] &  $95 \pm 10$ & $98 \pm 10$ \\
  \vmac\  [\kms] &  $65 \pm 10$ & $65 \pm 10$ \\
  $v_{\mathrm{eq}}$ [\kms] & $107 \pm 15$& $110 \pm 15$ \\
  $M$ [\msun] & $39.5^{+5.4}_{-4.4}$  &  $33.5^{+4.6}_{-3.7}$   \\ [3pt]
  $M_{\mathrm{spec}}$ [\msun] & $37.7^{+29.2}_{-16.1}$  & $31.8^{+26.6}_{-14.4}$   \\ [3pt]
  \hline
\end{tabular}
\tablefoot{(He/H)$_{\odot} = 0.1$, (C/H)$_{\odot} = 2.5 \times~10^{-4}$, (N/H)$_{\odot} = 6.0 \times~10^{-5}$ and (O/H)$_{\odot} = 4.6 \times~10^{-4}$ in number. $v_{\mathrm{cl}}$ indicates how rapidly the clumping should be switched on. }
\end{table}

The individual parameters agree with two supergiant objects according to the tables of \citet{martins05}, and \citet{bouret12}. The best-fit is achieved with $\teff = 32000$~K and 30500~K and $\logg = 3.15$ and 3.10 for the primary and the secondary, respectively. We also derive an overabundance of helium and nitrogen as well as a depletion of carbon in both objects, as expected from their luminosity class. Furthermore, we emphasize that the oxygen is depleted for the primary whilst it is solar for the secondary.

The wind parameters are reminiscent of those of Galactic supergiant O-type stars, according to \citet{bouret12} even though stars as cool as the two components of HD\,166734 were not analyzed by these authors. These parameters are thus in agreement with those of single massive supergiant objects.

   \section{Wind-wind interactions}
   \label{sec:wind}
   The spectral disentangling has yielded the mean spectrum of each component. From these spectra, the emission observed in the \ion{He}{ii}~4686 and the H$\alpha$ lines seems to be linked to the primary component over the entire orbital cycle. Our CMFGEN analysis shows that this component has a larger mass-loss rate than the secondary, whilst its terminal velocity is assumed to be similar.

   \begin{figure}
     \centering
     \includegraphics[width=9cm,bb=23 5 532 404,clip]{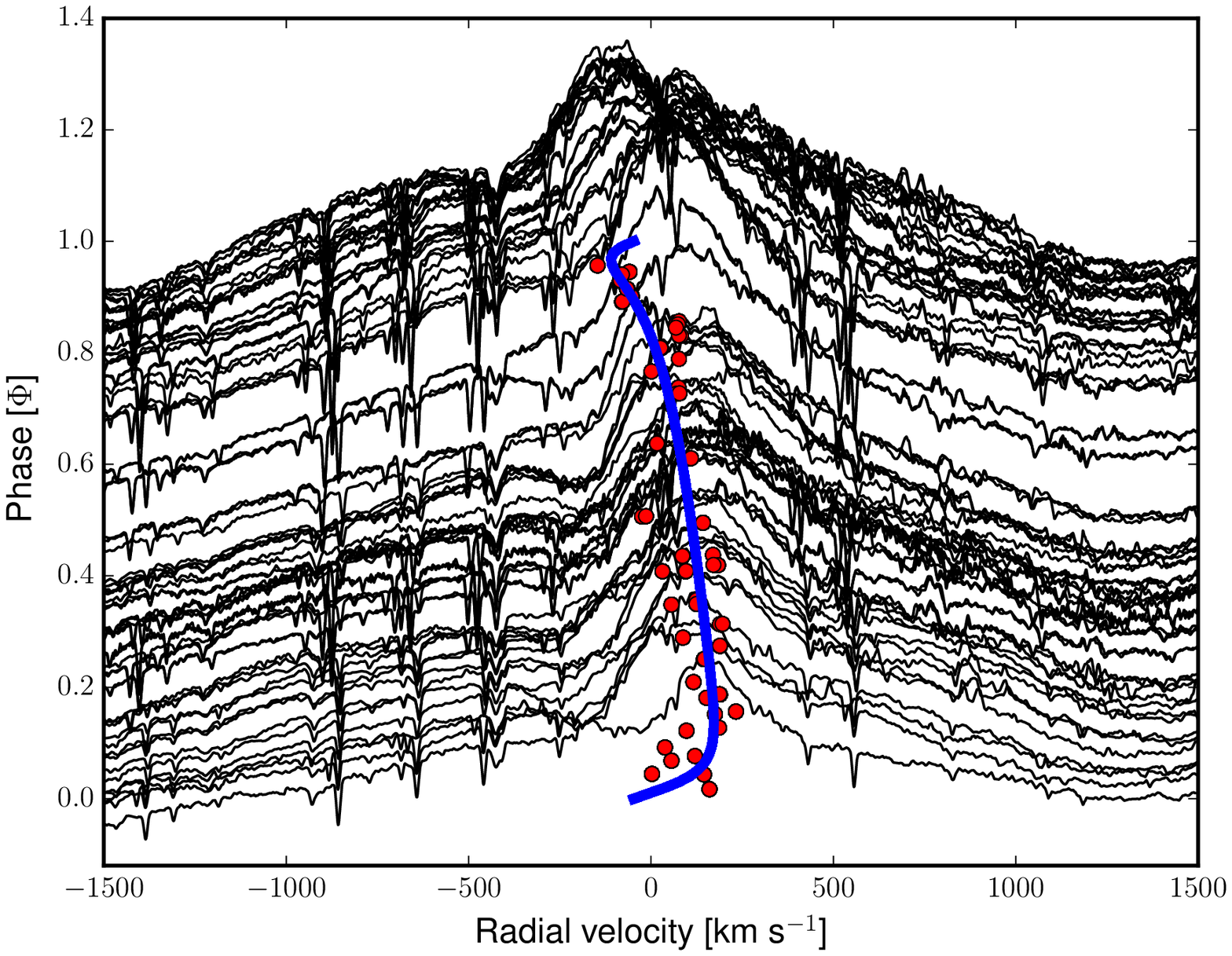}
     \includegraphics[width=9cm,bb=23 5 532 404,clip]{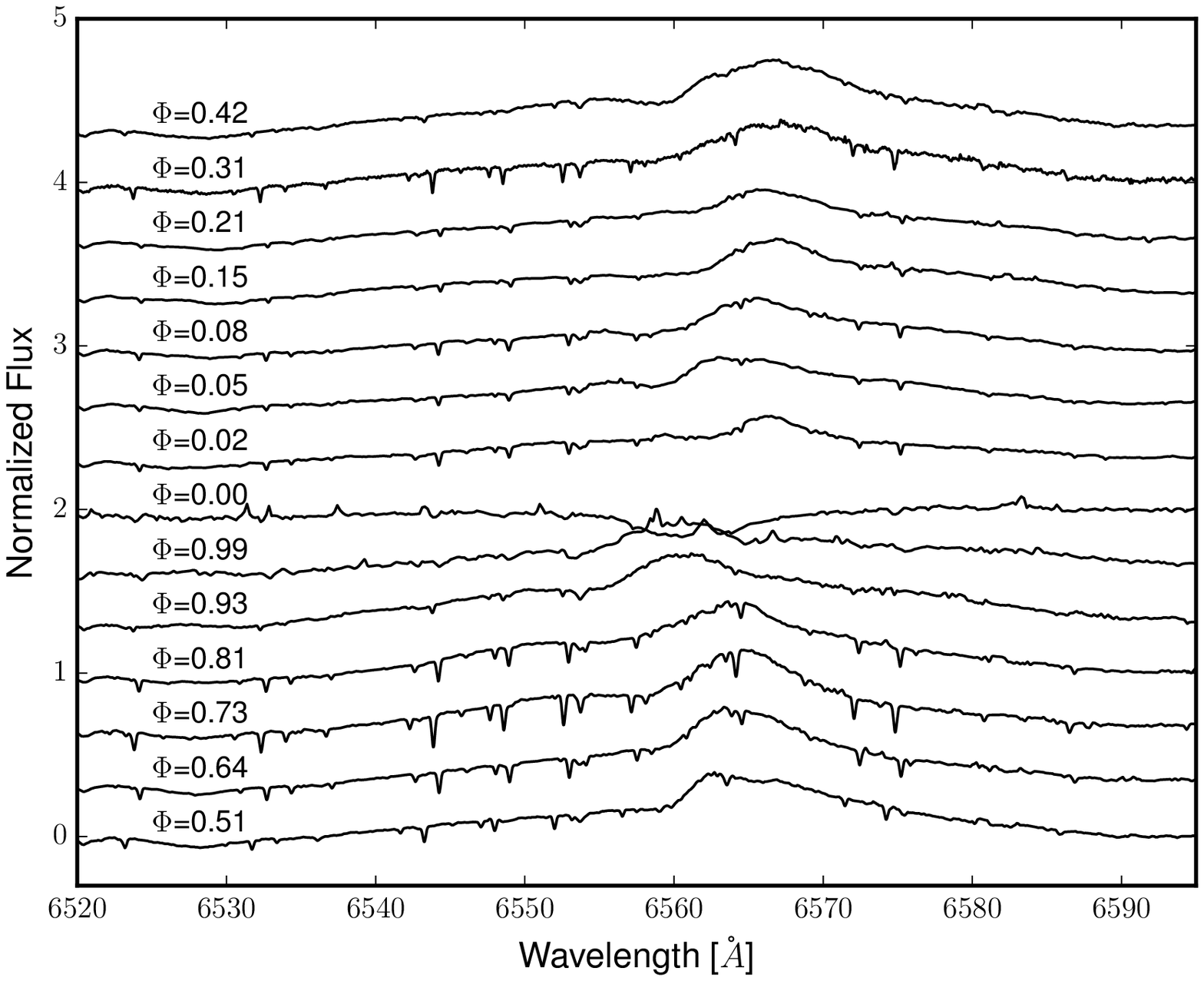}
     \caption{{\it Top:} Positions of the maximum of H$\alpha$ emission as a function of the orbital phase. The red dots represent the positions of the maximum, whilst the blue line is the radial velocity curve of the primary component. {\it Bottom: } Spectral variation of the H$\alpha$ line as a function of the orbital phase. The emission vanishes at $\Phi=0.00$.}\label{fig:wind} 
   \end{figure}

   Upon inspection of the H$\alpha$ profile in the observed spectra, it is clear that the maximum of emission of this line is associated to the primary on the whole orbit (see the upper panel of Fig.\,\ref{fig:wind}). A closer look at this profile at different phases (see the lower panel of Fig.\,\ref{fig:wind}) shows that at the periastron ($\Phi=0.0$) no emission is observed in the spectra, but this emission is present at $\Phi \sim 0.02$ and even at $\Phi \sim 0.99$. During this short time interval, when the two components are closest to each other, it appears that the emission generated by the two winds vanishes.

   The configuration of the system indicates that at the periastron the distance between the centers of both components is 71.2\,\rsun. Therefore, the distance that remains between the two photospheres (of about 16.9\,\rsun) is not large enough for the winds to strongly collide.
   
   \section{Discussion}
   \label{sec:discussion}
   
   \subsection{Stellar evolution}
   
   The stellar luminosities determined from photometry as well as the effective temperatures derived from the CMFGEN modeling provide the exact locations of both stars in the Hertzsprung-Russell diagram. They are displayed in Fig\,\ref{fig:HR}. Through our analysis, the orbital parameters of HD\,166734 as well as the stellar parameters and the derived surface abundances of its two components suggest that they have evolved mainly as single stars (at least without Roche lobe overflow event). They are thus interesting objects to be compared with the current evolutionary tracks for massive stars.

     The input physics and its implementation in six different evolutionary codes as well as their comparison with observed data were deeply discussed by \citet{martins13}. In the present discussion, we compare the observational parameters of the two components of HD\,166734 with the tracks of Geneva \citep{ekstrom12} and with those of \citet[][referred as STERN models by \citealt{martins13} and hereafter in this paper]{brott11}.
   \subsubsection{Geneva models}
   The evolutionary tracks of \citet{ekstrom12} are computed with a rotational rate $v_{\mathrm{eq}}/v_{\mathrm{crit}}$ of 0.4, a metallicity of $Z=0.014$, an overshoot parameter of 0.1 and do not account for magnetism. The upper panel of Fig.\,\ref{fig:HR} displays the evolutionary tracks and the isochrones of Geneva computed with rotation. Given the error bars on the luminosity and on the effective temperature, the two components are located, according to these tracks, close to the characteristic hook at the end of the main sequence. These positions therefore infer two different values for the initial mass of the two components. Indeed, within the error bars, the stars can be located either at the end of the main sequence, immediately before the stage where their temperature starts to increase, or before the yellow supergiant phase. Depending on their exact locations, the initial masses will be expected to be 44 or 47\,\msun\ for the primary and 37 or 39\,\msun\ for the secondary. According to the theoretical predictions, these stars would have ages between 5.2 and 5.4 Myrs for the primary and between 5.8 and 6.0 Myrs for the secondary (depending on the locations of both stars on the hook at the end of the main sequence). Furthermore, these tracks yield a radius of 27\,\rsun\ for the primary, and of 26\,\rsun\ for the secondary.

   We also determine the present evolutionary masses of each component. For the primary we compute a mass of $34.5^{+10.2}_{-9.5}$\,\msun\ whilst for the secondary we obtain a mass of $31.0^{+10.5}_{-9.5}$\,\msun. These masses can be compared to the spectroscopic masses computed from the luminosity, the effective temperature and the gravity of each star to be $37.7^{+29.2}_{-16.1}$\,\msun\ for the primary and $31.8^{+26.6}_{-14.4}$\,\msun\ for the secondary as well as to the dynamical masses of $39.5^{+5.4}_{-4.4}$\,\msun\ and $33.5^{+4.6}_{-3.7}$\,\msun, for the primary and the secondary, respectively. The large rotational rate ($v/v_{\mathrm{crit}}=0.4$) is however not realistic for massive stars in the Milky Way \citep{simondiaz14}. This infers an overestimation of the ages of the stars, and an underestimation of their evolutionary masses. From the tracks of Geneva, the two components of HD\,166734 are thus located close to the terminal age main sequence (TAMS) but given the fast evolution at the end of the main sequence, this part of the HR diagram is expected to be empty \citep{mcevoy15}. The inclusion of extra mixing (either by tidal effects at periastron or by increasing the overshoot parameter) in the evolutionary models would allow to increase the lifetime of the stars on the main sequence. Even though the initial rotational rate is large for the Geneva models, the spin down is quick. Indeed, for our individual stars taken as single, the actual equatorial velocities are predicted to be 15\,\kms\ for the primary and 18\,\kms\ for the secondary, which is too small with respect to the rotational velocities of our stars. The spin down is notably ruled by the mass-loss rate included in the models, suggesting that the predictions concerning the mass-loss in the Geneva models seems to be overestimated or that another phenomenon is at work.

   The initial rotational velocity of about 300\,\kms\ seems therefore too large for both components. We thus quickly compare the positions of the stars to the Geneva evolutionary tracks computed without rotation. From these models, the stars are on the main sequence, with estimated ages of about 3.0 Myrs for the primary and 3.4 Myrs for the secondary. The primary has an initial mass of 59\,\msun\ and an actual mass of 51.4\,\msun\ whilst the secondary has an initial mass of 50\,\msun\ and an actual mass of 45.3\,\msun; thus larger than what we found for the two components from evolutionary tracks with rotation as expected from the theory.

   \subsubsection{STERN models}

\begin{figure}
     \centering
     \includegraphics[width=8cm,bb=23 8 526 398,clip]{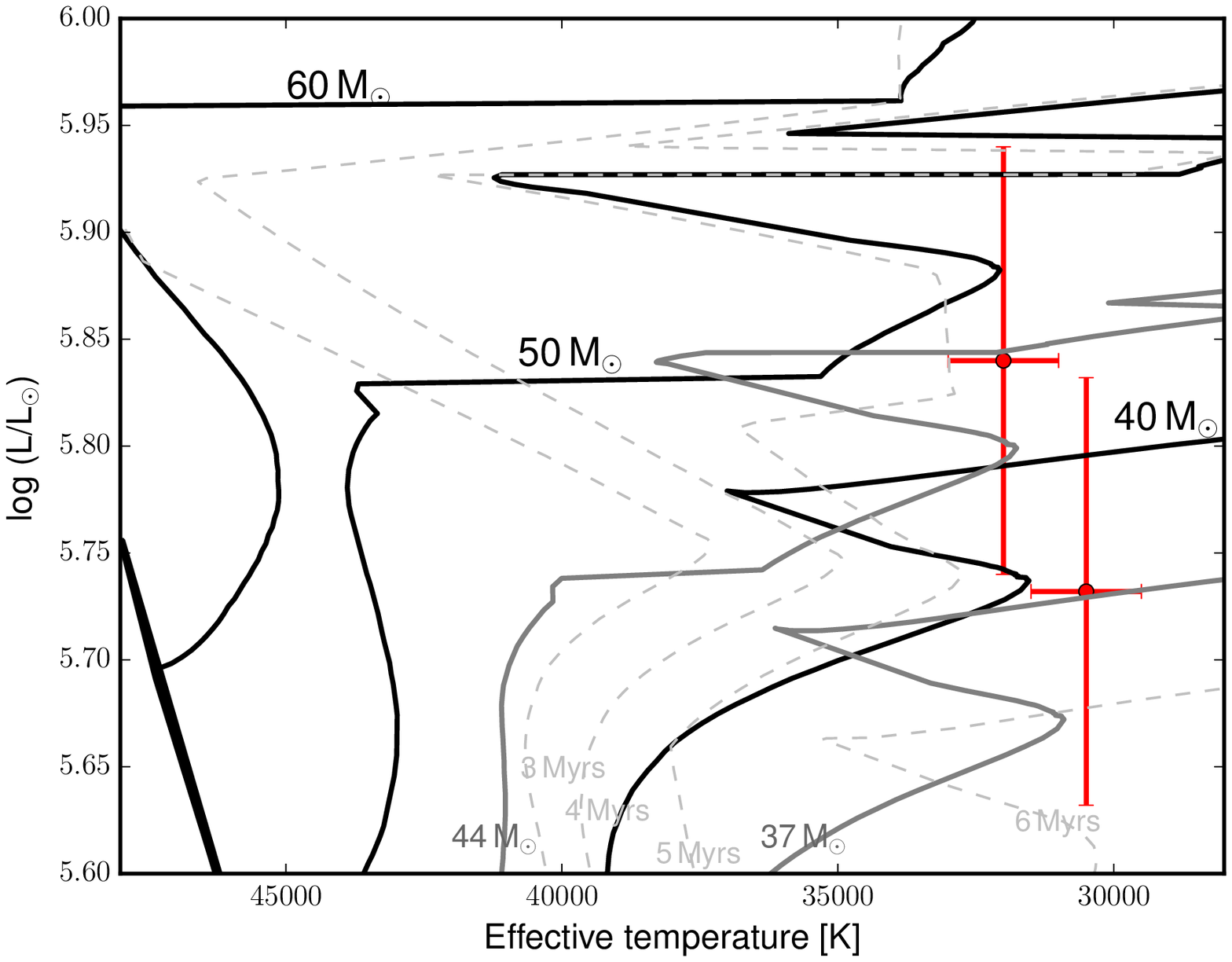}
     \includegraphics[width=8cm,bb=23 8 526 398,clip]{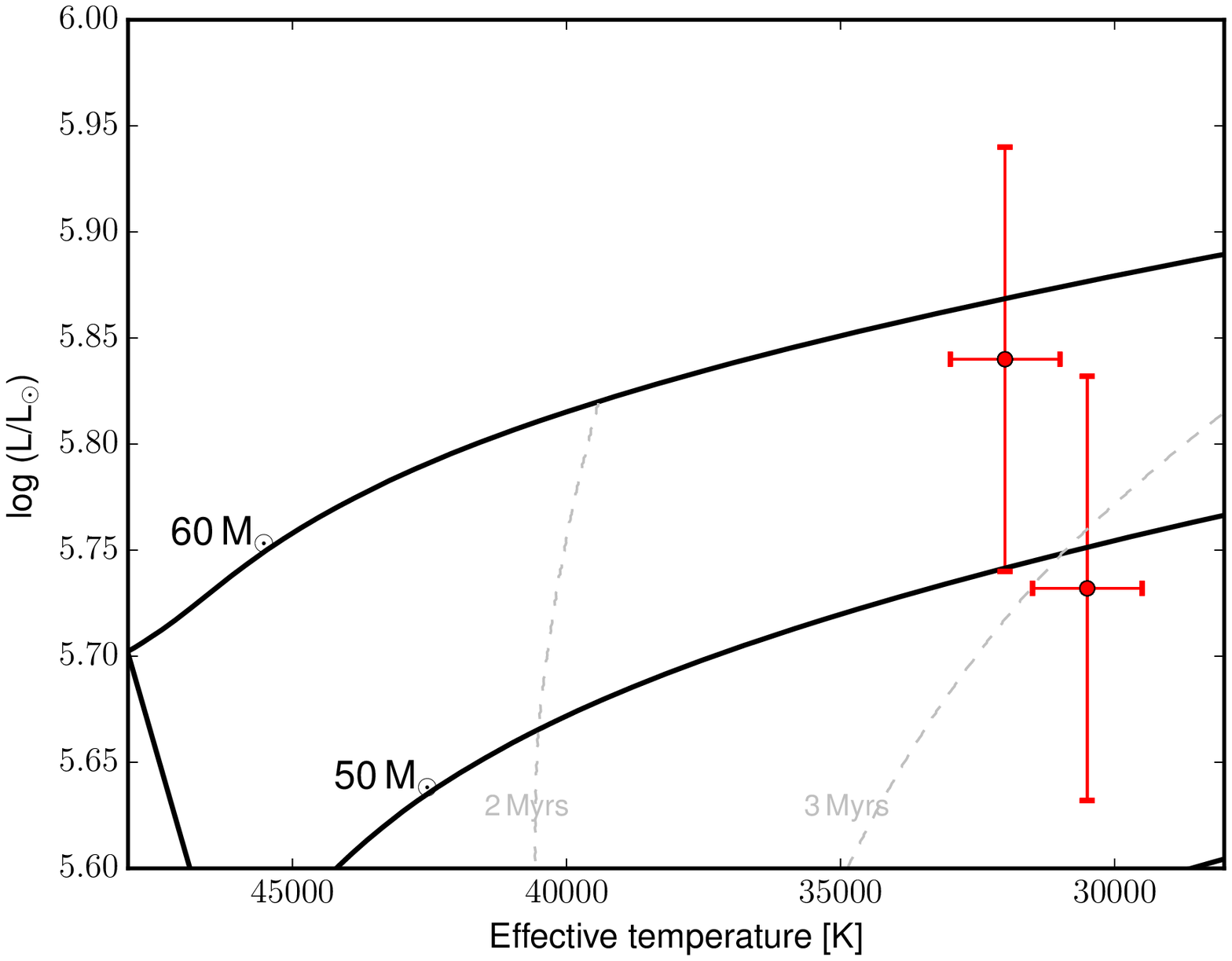}
     \caption{Positions in the HR diagram of the two components of HD\,166734. {\it Top: }Evolutionary tracks (black lines) and isochrones (gray lines) from \citet{ekstrom12}, computed with a rotational rate of 300\,\kms. {\it Bottom: }Evolutionary tracks (black lines) and isochrones (gray lines) from \citet{brott11}, computed with a rotational rate of 160\,\kms. The error bars represent the 1-$\sigma$.}\label{fig:HR} 
   \end{figure}

  We decided to compare our observational values with the STERN evolutionary models \citep{brott11}. These models constitute a grid with a sufficiently fine mesh, attenuating the errors due to the interpolation. This contrasts with the Geneva models for which one rotational rate is taken into account. The STERN models are computed with a metallicity of $Z=0.0088$, an overshoot parameter of 0.335 and the rotation is taken as a free parameter. The BONNSAI facility\footnote{The  BONNSAI  web-service  is  available  at http://www.astro.uni-bonn.de/stars/bonnsai} \citep{schneider14} was used as a Bayesian approach to derive the distribution of stellar parameters for each component of HD\,166734. The best fits provide an initial rotational velocity of 150\,\kms\ for both objects, giving an equatorial rotational velocity of 110\,\kms\ at their current stage (in adequation with our observational parameters). The lower panel of Fig.\,\ref{fig:HR} displays the locations of the stars on the HR diagram build from the STERN evolutionary tracks and the isochrones computed with an initial rotational velocity of 160\,\kms. For the primary, the initial mass is expected to be 56.1$_{-5.4}^{+4.7}$\,\msun, giving an evolutionary mass of 47.8$_{-4.4}^{+5.7}$\,\msun. For the secondary, the initial mass is 47.4$_{-4.6}^{+4.0}$\,\msun and the evolutionary mass is 41.2$_{-3.5}^{+4.7}$\,\msun. The STERN models provide an age of about 3.0\,Myrs for the primary and of about 3.5\,Myrs for the secondary; in agreement with the non-rotating tracks of Geneva. As emphasized by \citet{martins13}, the characteristic hook at the end of the main sequence occurs at lower temperatures because of the large amount of overshooting. These authors also mentioned that an overshooting parameter of 0.335, as used for the computation of the STERN models, is too large for stars with masses above 20\,\msun. Therefore, the ages estimated from these models are probably underestimated and their evolutionary masses overestimated. However, the STERN tracks better reproduce the rotational velocities of the supergiants. 

   \subsubsection{Surface abundances}
   
\begin{figure}
     \centering
     \includegraphics[width=8cm,bb=15 8 526 398,clip]{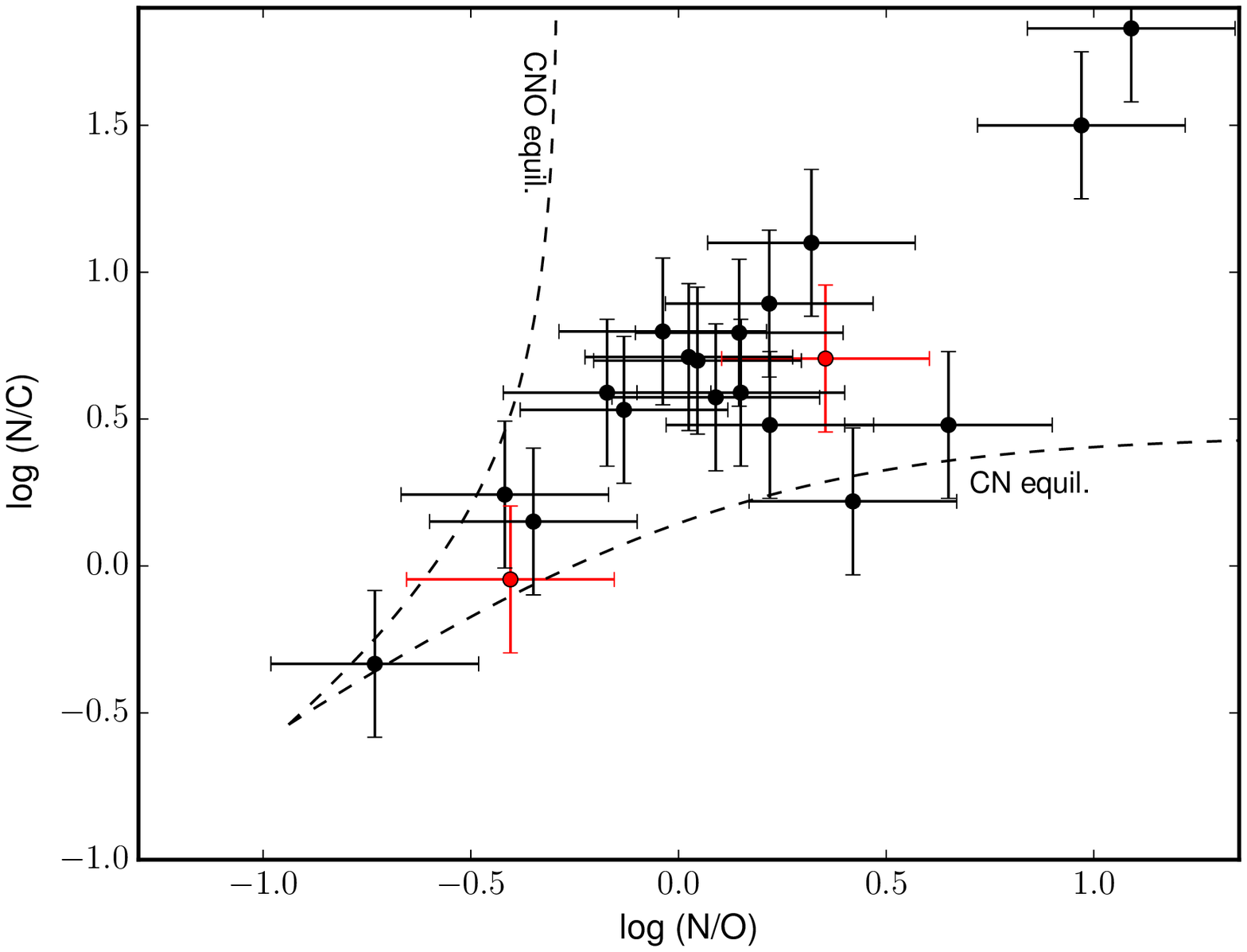}
     \includegraphics[width=8cm,bb=15 8 526 398,clip]{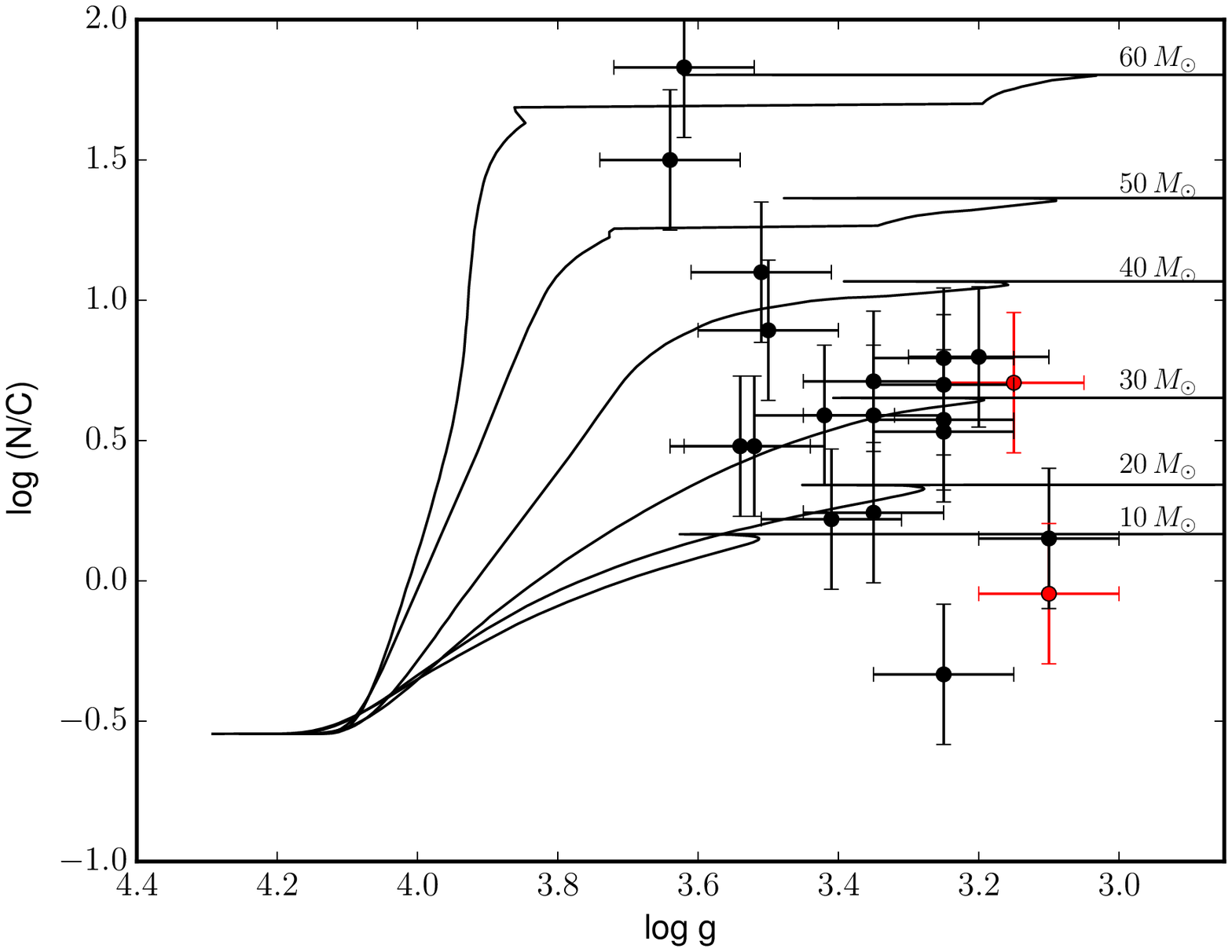}
     \caption{{\it Top: } $\log$ (N/C) (by number) as a function of $\log$ (N/O). The expected trends for the case of CN and CNO equilibrium are shown by the dashed lines. {\it Bottom:} $\log$ (N/C) (by number) as a function of \logg. The two components of HD\,166734 are represented by red dots, whilst the sample of Galactic supergiant stars analyzed by \citet{bouret12} and \citet{martins15} is represented by black dots. The evolutionary tracks are from \citet{ekstrom12} and are computed for a rotational rate (initial over critical) of 0.4. We use evolutionary tracks with initial masses larger than 10\,\msun.}\label{fig:abundances} 
   \end{figure}

   Whatever the evolutionary tracks we use, the secondary component appears more evolved than the primary. An enhanced mixing through tides at periastron could explain this difference in age (see Sect.\,\ref{synchro}). As we already mentioned in Section\,\ref{sec:modeling}, the primary and the secondary exhibit an overabundance in helium and in nitrogen as well as a depletion in carbon. Furthermore, the primary is also depleted in oxygen. The abundances thus confirm the evolved status of each object. However, they do not differ from abundances of single evolved massive stars. When we compare the surface abundances of the two components of HD\,166734 to the values of only the supergiant stars determined by \citet{martins15} in the framework of the MiMeS project and to the values derived by \citet{bouret12} in their analysis of a sample of Galactic supergiant stars, we see a clear correlation between N/O and N/C (see the upper panel of Fig.\,\ref{fig:abundances}) that does not differ from the results of \citet{bouret12} or of \citet{martins15}, represented in the upper panel of Fig.\,\ref{fig:abundances} by black dots. This agrees with the expectations of nucleosynthesis through the CNO cycle. The ratios N/C and N/O follow those of single stars according to the evolutionary models, within the uncertainties. The lower panel of Fig.\,\ref{fig:abundances} shows that the ratio N/C increases as surface gravity decreases, as was observed for single supergiant Galactic massive stars \citep{bouret12,martins15}. The ratios of N/C and N/O determined from the two components of HD\,166734 are thus correlated and in the same range as those measured for presumably single supergiant objects, meaning that no or at least few interactions occur between these two components. 

   \subsection{Synchronization of the system}
   \label{synchro}
   HD\,166734 is a highly-eccentric binary system where the secondary appears to be more evolved than the primary. This problem linked to the ages can be due to tidal mixing at periastron. This tidal mixing would affect the secondary more than the primary because of its size. In addition, some matter could be transferred to the primary with the consequence of increasing its spin, inducing more mixing and therefore a higher temperature; this increase of the temperature would modify the apparent age of the primary by becoming younger. This effect should be however limited because of the configuration of the system.

     We then may wonder whether these tidal interactions (that would be responsible for this tidal mixing) can affect the rotations of the components and can possibly synchronize the system (tidal interactions can indeed be responsible for a long timescale evolution of the binary orbital elements and the stellar rotation). To characterize these interactions in HD\,166734, we focus on the theory of \citet{hut81,hut82}. If we compute the ratio between the rotational and the orbital angular momentum (Eq.~14 in \citealt{hut81}), we find for the present stars of HD\,166734 that a large part of the angular momentum is contained in the orbit. An important classification parameter ($\alpha$) representing the ratio of the orbital angular momentum to the rotational angular momentum at equilibrium was introduced by \citet{hut81}. According to the value of this parameter, the circularization will take much more time than the synchronization. In the case of HD\,166734, for which the orbital separation between the two components is relatively large, the $\alpha$ parameter is extremely large, meaning that the circularization time is expected to be longer than the synchronization time. Always according to Hut's theory, the eccentricity of the orbit should decrease very slowly as well as the semimajor axis ($de/dt$ and $da/dt <0$) whilst the rotational angular velocities of both stars should increase as a function of time ($d\Omega/dt > 0$) but all this evolution is expected on timescales longer than the current age of the system. In eccentric binaries, the term of synchronization between the rotation and the revolution is rather replaced by the pseudo-synchronization at the moment of the periastron. This pseudo-synchronization of the system is reached when the rotational angular velocity of a star is at $0.8$ times the instantaneous orbital angular velocity at periastron \citep[see][]{hut81}. Indeed, assuming that the rotational axes of both components are perpendicular to the orbital plane, we use the combination of the inclination ($63.0\degr \pm 2.7\degr$) of the system, the equatorial rotational velocities ($107$ and $110$\,\kms) and the radii ($27.5$ and $26.8$\,\rsun) of both components, obtained in Section\,\ref{photometry}, to compute rotational periods of 13.0 days for the primary and of 12.3 days for the secondary. Both stars appear to have their rotations almost synchronized with each other but these values are smaller than the orbital period. When we compare the rotational angular velocities with the instantaneous orbital angular velocity at periastron, we compute ratios of about 0.6 for the primary and secondary components. We therefore conclude that pseudo-synchronization \citep{hut81} has not yet been achieved for HD\,166734. 
   
   \subsection{Distance to HD\,166734}

   In Section\,\ref{photometry}, we determined the bolometric magnitudes of $-9.85 \pm 0.23$ for the primary and $-9.58 \pm 0.26$ for the secondary, giving stellar luminosities of $\lL = 5.840 \pm 0.092$ and $\lL = 5.732 \pm 0.104$, respectively. Given their spectral types and their effective temperatures, we can compute, from \citet{martins06}, bolometric corrections of $-2.97 \pm 0.09$ and $-2.83 \pm 0.10$ for the primary and the secondary, respectively. From these values, we compute absolute magnitudes of $M_V = -6.88$ for the primary and $M_V = -6.75$ for the secondary.

   From the observed $V$ magnitude of HD\,166734 of 8.42 and its $B-V$ of 1.07 \citep{ducati02}, accounting for $(B-V)_0 = -0.27$ \citep{martins06}, and a brightness ratio of 1.28 between the primary and the secondary components, we estimate, for HD\,166734, a distance modulus of 11.85 and thus a distance of $2.34 \pm 0.30$\,kpc, confirming the distance estimated by \citet{conti80}.

   \begin{figure}
     \centering
     \includegraphics[width=8cm,bb=23 0 526 398,clip]{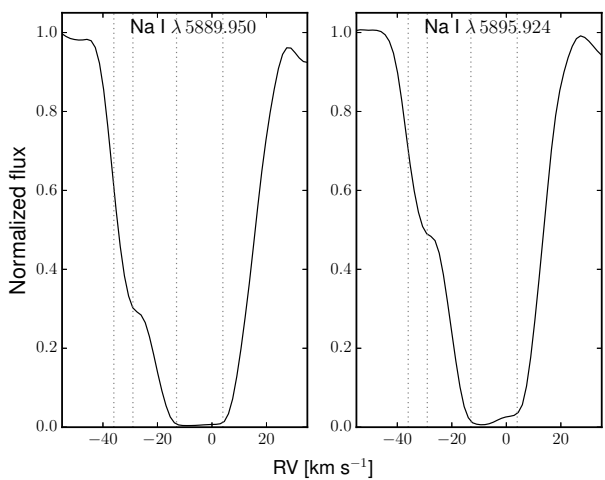}
     \caption{Section of the FEROS spectrum of HD\,166734 showing the \ion{Na}{i} interstellar lines, covering a velocity range of $-55$ to $+35$\,\kms. The dotted lines indicate, from left to right, RVs of $-36$, $-29$, $-13$, $+4$\,\kms\ for both panels.}\label{fig:NaI} 
   \end{figure}

   Next, we compare this distance with the kinematic distance estimated from the Galactic rotation curve. In Fig.\,\ref{fig:NaI}, we show the \ion{Na}{i} absorption line profiles. The \ion{Na}{i} lines are relatively saturated on a width ranging from $+4$ to $-13$\,\kms\ and with a shallower absorption profile between $-29$ and $-36$\,\kms. From the Galactic rotation model of \citet{fich89}, this velocity range gives a mean distance of $2.38 \pm 0.17$~kpc. This distance is in agreement with the distances estimated from the stellar luminosities or derived by \citet{conti80}.


   \section{Conclusions}
   \label{sec:conclusion}

   We present the results of the spectroscopic and photometric campaigns devoted to HD\,166734. The spectra taken close to the periastron passage have allowed us to refine the orbital solution of the system, removing the existing doubts on the masses of both components. Our analysis has indeed shown that HD\,166734 is a binary system with an orbital period of 34.537723 days and an eccentricity of 0.618. From the photometry, we determine an inclination of $63.0\degr\pm 2.7\degr$ for the system and actual masses of $39.5^{+5.4}_{-4.4}$\,\msun\ and $33.5^{+4.6}_{-3.7}$\,\msun\ for the primary and the secondary. 

   The individual spectra of each component, obtained by spectral disentangling, indicate a primary with a spectral classification of O7If, an effective temperature of $32000\pm1000$\,K, a \logg\ of $3.15\pm0.10$ and a luminosity of $\lL = 5.840 \pm 0.092$. We found that the secondary has the spectral type O9I(f) and its stellar parameters are $\teff = 30500\pm 1000$\,K, $\logg = 3.10\pm 0.10$, and $\lL = 5.732\pm 0.104$. We thus conclude, unlike \citet{conti80}, that the hottest and the most luminous star is also the most massive one. Both components show modified helium, carbon, nitrogen and oxygen surface abundances. They are both enriched in helium and nitrogen and are depleted in carbon. In addition, the primary is also depleted in oxygen. Their surface abundances do not however differ from the predictions of the `classical' evolutionary paths for single stars. This means that no or, at most, few interactions occur between the two stars, making them perfectly suitable as test objects for evolutionary models.
   
   Systems containing a pair of massive evolved O-type stars are relatively rare. Despite their number, they represent excellent tests for atmospheric and evolutionary models. Analyses of the kind we did for HD\,166734 are thus crucial to validate these models. In this context, we may quote other systems that can help with this validation. An interesting system is R139 \citep{taylor11} composed of an O6Iafc primary and of an O6Iaf secondary on a wide eccentric orbit. This system is the most massive binary system known to host two evolved Of supergiants but the physical parameters of each component are not complete enough to allow a deep comparison with the evolutionary models. Finally, WR21a \citep{tramper16, gosset17} is a very massive system composed of two younger O3/WN5 and O3 stars. Although this system is less evolved, it has orbital period ($P=31.67$\,days) and eccentricity ($e=0.64$) comparable to HD\,166734.

     In the present paper, we compare the fundamental parameters of each component with evolutionary tracks of Geneva \citep{ekstrom12} and of STERN \citep{brott11}. We conclude from our observations that the tracks of Geneva have an overly large initial rotational rate, that shortens the lifetime of the star on the main sequence and it seems that the mass-loss used to compute the tracks is also too large to reproduce the actual rotational velocities of the stars. At the same time, the STERN tracks have an overly large overshoot parameter, but the spin down of the stars is sufficient to agree with the actual rotational velocities of the stars. Finally, the ages and the evolutionary masses for both components are different as a function of the tracks used in the analysis. Even though other systems (such as those quoted above) must still be investigated, HD\,166734 allowed us to show that the evolutionary models are not reliable enough to reproduce the exact characteristics of supergiant stars.

   \appendix

   \begin{acknowledgements}
     The authors thank the anonymous referee for his/her constructive remarks on the paper. This research was supported by the Fonds National de la Recherche Scientifique (F.R.S.-F.N.R.S.), by the PRODEX XMM contract (Belspo),  and through the ARC grant for Concerted Research Actions, financed by the French Community of Belgium (Wallonia-Brussels Federation). L.M. thanks Pavel Mayer for his advice in the light curve modeling and John Hillier for making CMFGEN available. We are grateful to the staff of La Silla ESO Observatory for their technical support.
   \end{acknowledgements}

   \bibliography{hd166734.bib}

   \begin{appendix}
     \section{Journal of the spectroscopic observations of HD\,166734}
     \begin{table*}
       \caption{Journal of the observations of HD\,166734.}\label{tab:journal}             
       \centering                          
       \begin{tabular}{l c r r r | l c r r r }        
         \hline\hline                 
         HJD--2\,450\,000 & Instrument & $\phi$ & RV$_{\mathrm{P}}$ &  RV$_{\mathrm{S}}$ & HJD--2\,450\,000 & Instrument & $\phi$ & RV$_{\mathrm{P}}$ &  RV$_{\mathrm{S}}$ \\
         \hline                        
2201.6084 & ESPRESSO & 0.189 &  94.7 & -102.5 & 3643.5890 &    FEROS & 0.940 & -186.1 & 199.6 \\ 
2201.6311 & ESPRESSO & 0.190 &  91.0 & -103.0 & 3643.6064 &    FEROS & 0.941 & -190.1 & 198.6 \\ 
2201.6636 & ESPRESSO & 0.191 &  93.3 & -101.2 & 3797.9001 &    FEROS & 0.408 &  52.4 & -44.3 \\ 
2202.5888 & ESPRESSO & 0.218 &  85.9 & -88.3 & 3800.9080 &    FEROS & 0.495 &  37.3 & -29.4 \\ 
2202.6127 & ESPRESSO & 0.219 &  84.4 & -86.2 & 3862.8646 &    FEROS & 0.289 &  57.2 & -85.5 \\ 
2202.6560 & ESPRESSO & 0.220 &  79.5 & -85.4 & 3864.9075 &    FEROS & 0.348 &  43.5 & -83.1 \\ 
2381.9031 &    FEROS & 0.410 &  41.1 & -63.3 & 3867.9057 &    FEROS & 0.435 &  37.7 & -30.5 \\ 
2381.9212 &    FEROS & 0.410 &  35.3 & -60.3 & 3867.9232 &    FEROS & 0.436 &  37.3 & -24.2 \\ 
2382.9269 &    FEROS & 0.439 &  37.8 & -59.8 & 3874.8779 &    FEROS & 0.637 &  12.1 &  -7.2 \\ 
2782.8462 &    FEROS & 0.019 & -67.2 &  77.3 & 3874.8953 &    FEROS & 0.638 &   7.2 &   9.1 \\ 
2783.7920 &    FEROS & 0.046 &  47.8 & -76.0 & 3880.7990 &    FEROS & 0.809 & -75.1 &  95.1 \\ 
2784.9020 &    FEROS & 0.078 &  63.7 & -87.5 & 3880.8165 &    FEROS & 0.809 & -73.3 &  95.6 \\ 
3130.7969 &    FEROS & 0.093 &  88.9 & -94.8 & 3899.7138 &    FEROS & 0.356 &  42.7 & -79.7 \\ 
3131.8345 &    FEROS & 0.123 &  93.4 & -118.2 & 3899.7312 &    FEROS & 0.357 &  45.4 & -82.3 \\ 
3132.8529 &    FEROS & 0.153 &  89.7 & -103.5 & 3912.8945 &    FEROS & 0.738 & -35.5 &  49.4 \\ 
3133.8712 &    FEROS & 0.182 &  89.1 & -108.5 & 3913.8940 &    FEROS & 0.767 & -56.9 &  62.5 \\ 
3134.8491 &    FEROS & 0.210 &  82.8 & -79.7 & 3914.6723 &    FEROS & 0.789 & -59.0 &  67.9 \\ 
3288.6241 & ESPRESSO & 0.663 &  -5.1 &   6.6 & 3918.5369 &    FEROS & 0.901 & -149.4 & 157.2 \\ 
3288.6393 & ESPRESSO & 0.663 &  -4.8 &   9.6 & 3918.5543 &    FEROS & 0.902 & -152.6 & 157.1 \\ 
3288.6539 & ESPRESSO & 0.664 &   2.1 &  -3.2 & 3919.5103 &    FEROS & 0.929 & -179.5 & 191.4 \\ 
3289.6186 & ESPRESSO & 0.692 & -31.3 &  72.1 & 3919.5277 &    FEROS & 0.930 & -179.9 & 193.2 \\ 
3289.6364 & ESPRESSO & 0.692 & -37.8 &  58.5 & 3950.6415 &    FEROS & 0.831 & -88.7 &  95.8 \\ 
3289.6513 & ESPRESSO & 0.693 & -26.6 &  54.6 & 3950.6589 &    FEROS & 0.831 & -87.4 &  97.7 \\ 
3291.6744 & ESPRESSO & 0.751 & -43.7 &  71.7 & 3951.5085 &    FEROS & 0.856 & -107.5 & 117.9 \\ 
3291.6890 & ESPRESSO & 0.752 & -49.0 &  63.1 & 3951.5304 &    FEROS & 0.856 & -106.5 & 117.4 \\ 
3291.7036 & ESPRESSO & 0.752 & -45.9 &  63.3 & 3952.7421 &    FEROS & 0.892 & -137.8 & 157.9 \\ 
3509.8746 &    FEROS & 0.069 &  49.8 & -82.9 & 3952.7595 &    FEROS & 0.892 & -140.9 & 159.6 \\ 
3511.9189 &    FEROS & 0.128 &  85.5 & -113.3 & 3953.5369 &    FEROS & 0.915 & -160.9 & 182.8 \\ 
3512.8354 & ESPRESSO & 0.155 &  81.2 & -90.9 & 3953.5543 &    FEROS & 0.915 & -161.4 & 181.6 \\ 
3512.8482 & ESPRESSO & 0.155 &  84.4 & -91.8 & 3954.5829 &    FEROS & 0.945 & -181.0 & 209.6 \\ 
3512.8653 & ESPRESSO & 0.155 &  85.3 & -90.6 & 3954.6004 &    FEROS & 0.945 & -181.1 & 210.9 \\ 
3512.8804 & ESPRESSO & 0.156 &  86.2 & -94.8 & 4193.9278 & ESPRESSO & 0.875 & -113.6 & 154.3 \\ 
3512.9230 &    FEROS & 0.157 &  99.8 & -99.3 & 4193.9416 & ESPRESSO & 0.875 & -104.1 & 162.8 \\ 
3513.8269 & ESPRESSO & 0.183 &  76.2 & -102.5 & 4194.9639 & ESPRESSO & 0.905 & -138.2 & 177.0 \\ 
3513.8443 & ESPRESSO & 0.184 &  73.3 & -93.2 & 4194.9788 & ESPRESSO & 0.905 & -130.8 & 183.8 \\ 
3513.8590 & ESPRESSO & 0.184 &  74.7 & -85.4 & 4194.9951 & ESPRESSO & 0.906 & -134.6 & 181.1 \\ 
3515.8300 & ESPRESSO & 0.241 &  84.9 & -86.9 & 4196.9348 & ESPRESSO & 0.962 & -195.0 & 233.4 \\ 
3515.8472 & ESPRESSO & 0.242 &  62.6 & -84.9 & 4196.9570 & ESPRESSO & 0.963 & -195.7 & 232.2 \\ 
3515.8619 & ESPRESSO & 0.242 &  66.9 & -87.0 & 4196.9760 & ESPRESSO & 0.963 & -198.2 & 226.2 \\ 
3516.8296 & ESPRESSO & 0.270 &  61.4 & -81.5 & 4197.8627 & ESPRESSO & 0.989 & -180.8 & 198.8 \\ 
3516.8445 & ESPRESSO & 0.271 &  51.1 & -80.2 & 4197.8852 & ESPRESSO & 0.989 & -178.5 & 195.9 \\ 
3516.8592 & ESPRESSO & 0.271 &  50.9 & -82.1 & 4197.9099 & ESPRESSO & 0.990 & -176.1 & 193.4 \\ 
3517.8129 & ESPRESSO & 0.299 &  60.6 & -67.5 & 4199.8838 & ESPRESSO & 0.047 &  42.1 & -47.5 \\ 
3517.8275 & ESPRESSO & 0.299 &  60.6 & -73.6 & 4199.9010 & ESPRESSO & 0.048 &  38.2 & -49.8 \\ 
3517.8443 & ESPRESSO & 0.300 &  60.0 & -74.7 & 4199.9157 & ESPRESSO & 0.048 &  38.2 & -43.5 \\ 
3550.6987 &    FEROS & 0.251 &  75.5 & -99.5 & 4249.9649 & ESPRESSO & 0.497 &  26.2 & -17.9 \\ 
3550.7161 &    FEROS & 0.251 &  76.9 & -99.7 & 4249.9768 & ESPRESSO & 0.498 &  26.9 & -19.1 \\ 
3552.8692 &    FEROS & 0.314 &  56.7 & -70.2 & 4250.8564 & ESPRESSO & 0.523 &  24.5 & -30.7 \\ 
3552.8852 &    FEROS & 0.314 &  53.3 & -73.5 & 4250.8863 & ESPRESSO & 0.524 &  23.0 & -24.5 \\ 
3588.6468 &    FEROS & 0.350 &  51.6 & -69.6 & 4251.7829 & ESPRESSO & 0.550 &   7.9 &   0.1 \\ 
3588.6642 &    FEROS & 0.350 &  52.8 & -70.9 & 4251.8100 & ESPRESSO & 0.551 &   3.8 &   0.2 \\ 
3597.6673 &    FEROS & 0.611 &   4.9 &  -7.3 & 4251.8433 & ESPRESSO & 0.552 &   6.1 &   0.2 \\ 
3597.6847 &    FEROS & 0.611 &   6.1 &  -8.7 & 4982.8318 & ESPRESSO & 0.717 & -36.9 &  40.9 \\ 
3601.6977 &    FEROS & 0.727 & -24.2 &  32.2 & 4983.7593 & ESPRESSO & 0.743 & -11.0 &  81.1 \\ 
3601.7151 &    FEROS & 0.728 & -24.7 &  31.2 & 4984.7515 & ESPRESSO & 0.772 & -42.1 &  76.8 \\ 
3617.5842 &    FEROS & 0.187 &  87.8 & -96.2 & 4985.8245 & ESPRESSO & 0.803 & -59.3 &  97.0 \\ 
3617.6017 &    FEROS & 0.188 &  88.3 & -95.5 & 4986.8138 & ESPRESSO & 0.832 & -68.3 & 106.2 \\ 
3620.5870 &    FEROS & 0.274 &  57.8 & -88.1 & 4987.8062 & ESPRESSO & 0.861 & -109.4 & 141.2 \\ 
3620.6044 &    FEROS & 0.275 &  57.9 & -87.4 & 4988.7544 & ESPRESSO & 0.888 & -133.0 & 163.0 \\ 
3625.6001 &    FEROS & 0.420 &  41.9 & -70.1 & 4989.7606 & ESPRESSO & 0.917 & -160.6 & 186.2 \\ 
3625.6175 &    FEROS & 0.420 &  35.5 & -70.0 & 4992.7391 & ESPRESSO & 0.003 & -134.9 & 113.5 \\ 
3628.6180 &    FEROS & 0.507 &  25.0 & -12.0 & 5021.7712 &    FEROS & 0.844 & -88.0 & 110.3 \\ 
3628.6355 &    FEROS & 0.507 &  22.7 & -14.6 & 5025.6253 &    FEROS & 0.956 & -187.1 & 217.4 \\ 
3640.5021 &    FEROS & 0.851 & -105.4 & 123.6 & 5028.6212 &    FEROS & 0.042 &  15.9 &   3.5 \\ 
3640.5196 &    FEROS & 0.852 & -105.4 & 123.7 &           &          &       &       &       \\
         \hline                                   
       \end{tabular}
     \end{table*}

     \section{Periodograms}

   \begin{figure}
     \centering
     \includegraphics[width=9cm,bb=0 5 568 425,clip]{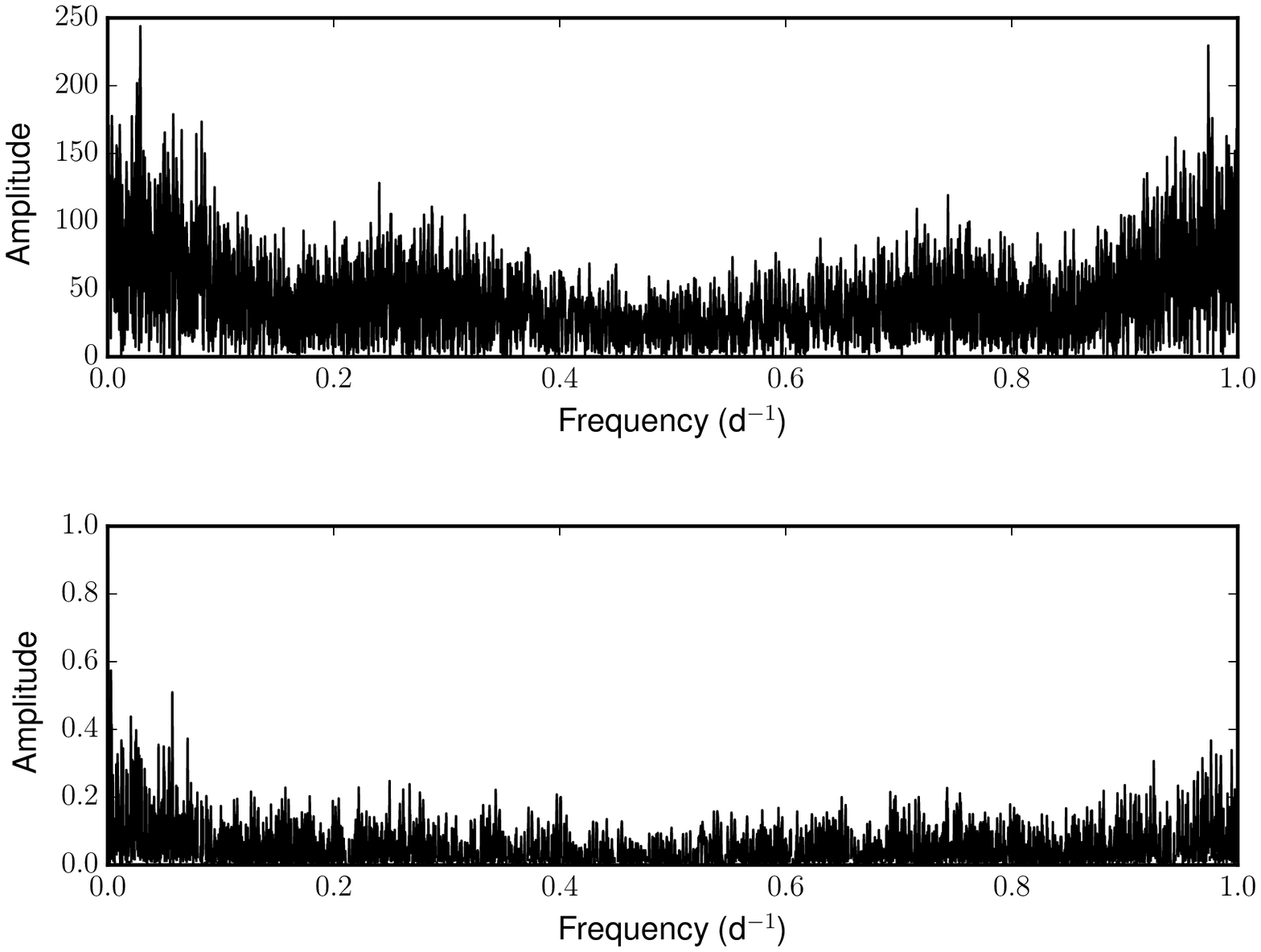}
     \caption{{\it Top:} Power spectrum computed on the basis of RV$_{\mathrm{S}}$--RV$_{\mathrm{P}}$. {\it Bottom:} Spectral window computed from the spectroscopic observations. }\label{fig:periodogramRV} 
   \end{figure}

   \begin{figure}
     \centering
     \includegraphics[width=9cm,bb=0 5 568 425,clip]{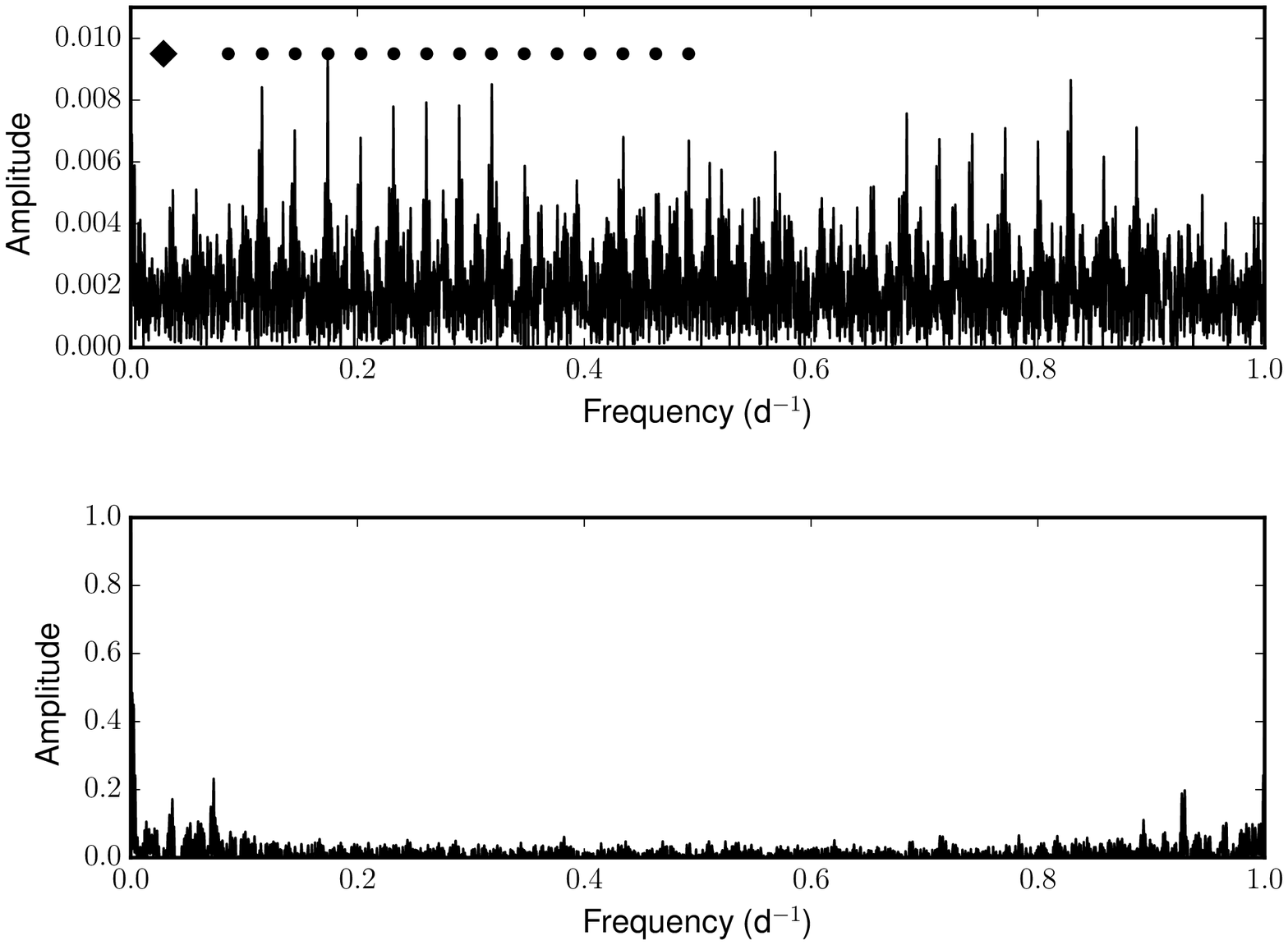}
     \caption{{\it Top:} Power spectrum of the {\it R} light curve of HD\,166734. The diamond indicates the location of the orbital frequency while the dots represent its harmonics. {\it Bottom:} Spectral window computed from the {\it R} light curve. }\label{fig:periodogramR} 
   \end{figure}

   \begin{figure}
     \centering
     \includegraphics[width=9cm,bb=0 5 568 425,clip]{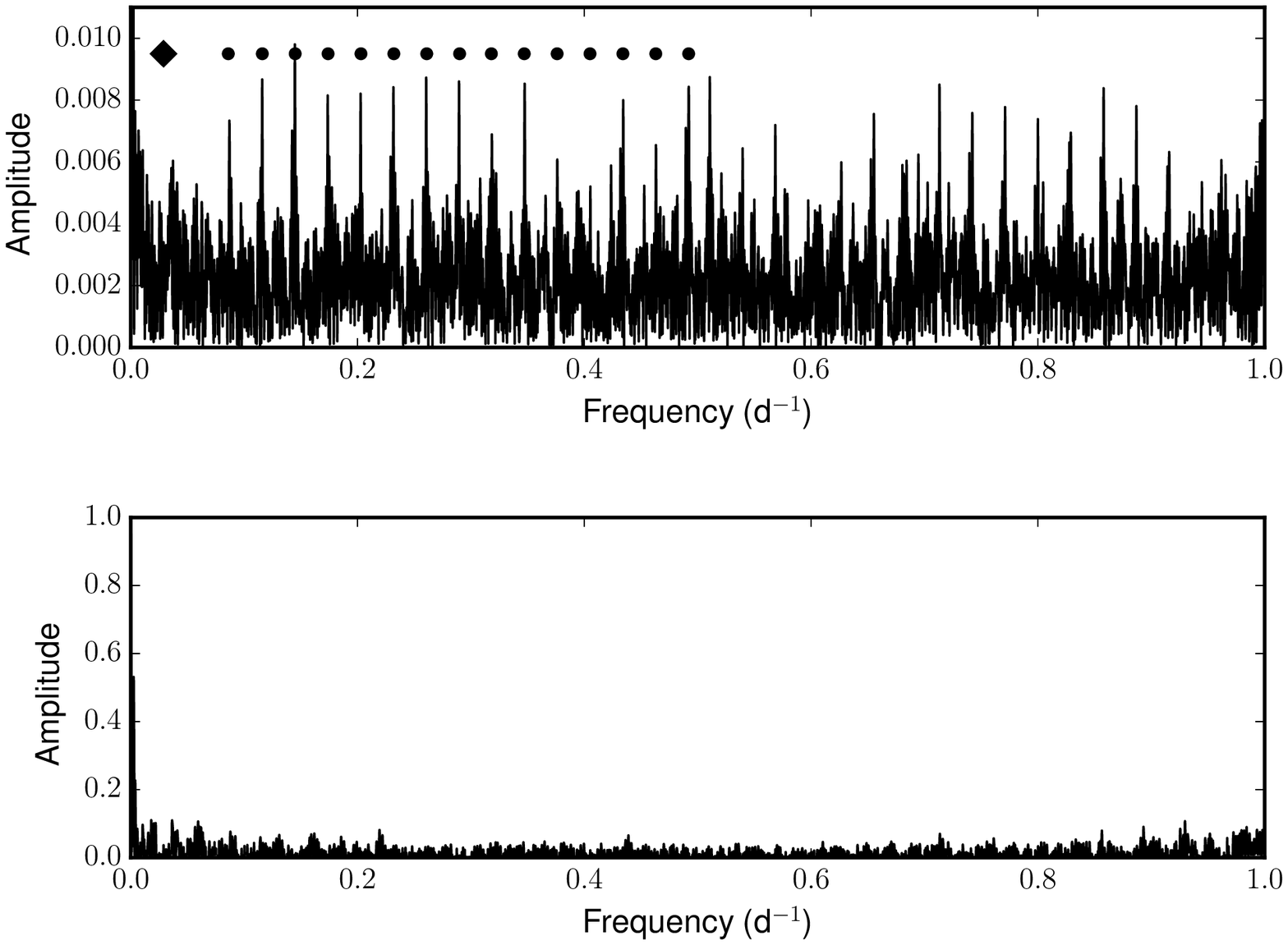}
     \caption{{\it Top:} Power spectrum of the {\it I} light curve of HD\,166734. The diamond indicates the location of the orbital frequency while the dots represent its harmonics. {\it Bottom:} Spectral window computed from the {\it I} light curve. }\label{fig:periodogramI} 
   \end{figure}

   \end{appendix}
\end{document}